\documentclass[twocolumn,preprintnumbers,amsmath,amssymb,pre]{revtex4-1}
\usepackage{natbib,hyperref}
\hypersetup{colorlinks=true, citecolor=blue, urlcolor=blue, linkcolor=blue}
\usepackage[english]{babel}
\usepackage[utf8]{inputenc}
\usepackage{graphicx}
\usepackage[caption=false]{subfig}
\usepackage{amsmath}
\usepackage{amsfonts}
\usepackage{amssymb}
\usepackage{color,soul}
\setulcolor{red}
\usepackage{xcolor}
\DeclareMathOperator\erf{erf}

\begin{document}

\title{Super-Arrhenius diffusion in a binary colloidal mixture at low volume 
fraction: An effect of depletion interaction due to an asymmetric barrier}

\author{Jalim Singh}
\author{Mahammad Mustakim}
\author{A. V. Anil Kumar}
\email{anil@niser.ac.in}
\affiliation{School of Physical Sciences, 
National Institute of Science Education and Research, HBNI,
Jatni, Bhubaneswar 752050, India.}

\date{\today}

\begin{abstract}

	We report results from the molecular dynamics simulations of a binary colloidal mixture
	subjected to an external potential barrier along one of the spatial directions at low volume
	fraction, $\phi$ = 0.2. The variations in the asymmetry of the external potential barrier do not
	change the dynamics of the smaller particles, showing Arrhenius diffusion. However, the
	dynamics of the larger particles shows a crossover from sub-Arrhenius to super-Arrhenius
	diffusion with the asymmetry in the external potential at the low temperatures and low volume
	fraction. Super-Arrhenius diffusion is generally observed in the high density systems where
	the transient cages are present due to dense packing, e.g., supercooled liquids, jammed
	systems, diffusion through porous membranes, dynamics within the cellular environment, etc.
	This model can be applied to study the molecular transport across cell membranes, nano-, and
	micro-channels which are characterized by spatially asymmetric potentials.

\end{abstract}

\maketitle

\section{Introduction}

Binary colloidal mixtures of unequal sizes serve as a paradigm for entropic
manipulation of structural as well as dynamical properties of soft
matter \cite{j:crocker,j:dickman,j:kaplan,j:eldridge,j:igwe}. The primary 
reason behind this is the attractive depletion interaction between the larger 
species of the mixture due to the presence of the smaller species. The physical 
interpretation of depletion interactions is that there exists an effective attractive 
interaction between the larger particles which favors the overlap of depletion layers
around them thereby providing a larger free volume available for the smaller 
particles \cite{lekkerkerker}. This interpretation hinges on the view that an osmotic
pressure imbalance pushes two larger particles close to each other when they reach within
a length scale set by the smaller particles. It was first proposed by Asakura and
Oosawa \cite{j:asakura} and later reconsidered by Vrij \cite{j:vrij} while explaining
the phase behavior of colloid-polymer mixtures. Depletion interactions are invoked to
explain a large variety of phenomena in mixtures such as colloidal
crystallization \cite{j:lin,j:kozina,j:hatch,j:oversteegen}, vitrification of 
colloids \cite{j:eckert,j:miller,j:dinsmore,j:williams,j:germain}, dynamics in crowded
medium \cite{j:toan,j:tuinier,j:ullmann}, etc.

It has been shown that in a binary colloidal mixture subjected to external 
potentials (symmetric barrier at the middle and confinement along the $z-$direction), 
there exists depletion interaction not only between the larger particles but also 
between the external potentials and the larger particles \cite{j:anil_sym_jcpstr}. When
the external potential is finite, this depletion interaction alters structural and 
dynamical properties significantly. When a binary mixture of colloids is subjected
to an external repulsive potential barrier, the depletion interaction between the
repulsive barrier and the larger particles leads to a demixing in the 
mixture \cite{j:anil_sym_jcpstr,j:roth}. Moreover, dynamical properties of both 
sizes of particles in the mixture deviate substantially in the presence of the 
external barrier (without the confining potential) and show a lot of 
interesting phenomena \cite{j:anil_sym_jcpdyn}. For example, at low temperatures,
the smaller particles get localized between the external potential barriers leading
to a slowing down in their dynamics, similar to the dynamics of supercooled 
liquids \cite{j:anil_sym_jcpdyn}. This slowing down is manifested by a plateau in
the mean-squared displacement at intermediate times, non-zero non-Gaussian parameter,
two-step relaxation in the self-intermediate scattering function, etc. This is intriguing
as this occurs even at a very low volume fraction. However, the larger particles do
not get localized between the barriers and continue to show normal diffusion even at 
low temperatures. This is attributed to the reduction in effective potential barrier
the larger particles have to cross during their dynamics due to the depletion
interaction. At high temperatures, the smaller particles diffuse faster than
the larger particles, which is expected. However, as temperature decreases, the 
larger particles start diffusing faster than their smaller counterparts, thus showing
a crossover in the diffusion coefficients. This occurs because the diffusion coefficient 
of the larger components ($D_l$) decreases very slowly with decreasing temperature.
This weak dependence of $D_l$ on temperature suggests that the diffusion of the
larger components is no longer Arrhenius and the activation energy for diffusion is
temperature dependent. It has been shown that the larger particles' diffusion is
sub-Arrhenius and activation energy decreases with temperature whereas the
diffusion of the smaller particles is Arrhenius with a constant activation
energy \cite{j:anil_soft_sym}. In general, sub-Arrhenius diffusion is considered 
to be intimately related to quantum phenomena and has been observed mostly in 
systems where quantum tunneling plays an important role such as in certain 
chemical reactions \cite{j:aquilanti,j:silva}. In fact, the binary colloidal
mixture, subjected to an external Gaussian potential barrier, is the first 
classical system which has been observed to undergo sub-Arrhenius 
diffusion \cite{j:anil_soft_sym}.

It is shown that an asymmetry of cell-membrane channel controls the transport 
across it by ratchet like mechanism, when non-equilibrium fluctuations are
present \cite{schulten}. Interestingly, Kolomeisky shows that asymmetry in the
external potential alters the molecular transport across cell membrane, even 
without nonequilibrium fluctuations \cite{kolomeisky}. This results in the 
asymmetric diffusion across such membrane channels in which particles' diffusion
dominates in one direction \cite{shaw}. It is interesting to study dynamical
properties associated with this anomalous diffusion. Here, we investigate the
effect of \textit{asymmetry} in the external potential barrier on the structure
and dynamics of the binary colloidal mixture, compared to 
Refs. \cite{j:anil_sym_jcpstr,j:anil_sym_jcpdyn} where the external potential 
barrier is \textit{symmetric}, using canonical ensemble molecular dynamics 
simulations. We find that as the asymmetry in the external potential increases,
the diffusion of the larger particles changes from sub-Arrhenius to super-Arrhenius.
This happens due to the crowding of the larger particles near the barrier. This 
crowding is more on the steeper side of the potential when its asymmetry
becomes large. A theoretical and experimental study on colloidal particles at
low volume fractions by Dalle-Ferrier \textit{et al.} also shows the dynamical
signatures of supercooled liquids in the system due to external sinusoidal 
potentials \cite{egelhaaf}. Interestingly, by calculating the self part 
of the van-Hove correlation function of small and large particles, along the 
applied external potential, we reveal that the transport of the 
particles is bidirectional. Thorn \textit{et al.} show the 
asymmetrical optical trapping of particles using two laser beams in which 
one laser beam produces the external potential such that atoms transmit to the 
other side, whereas the other laser beam represents a reflecting potential, 
thus demonstrate the unidirectional atomic motion \cite{thorn}. Our model can
be realized in such experiments where the external potential remains
asymmetric while maintaining bidirectional transport.

The remainder of this paper is organized
as follows: we begin with a description of our model and the simulation details
in Sec. \ref{s:model}. The dynamical properties as well as the structural changes 
associated with the asymmetric potential are presented in Sec. \ref{s:results}. 
Finally, we conclude our results in Sec. \ref{s:sumcon}.

\section{Model\label{s:model}}

\begin{figure}
	\includegraphics[width=7.5cm, height=5.5cm]{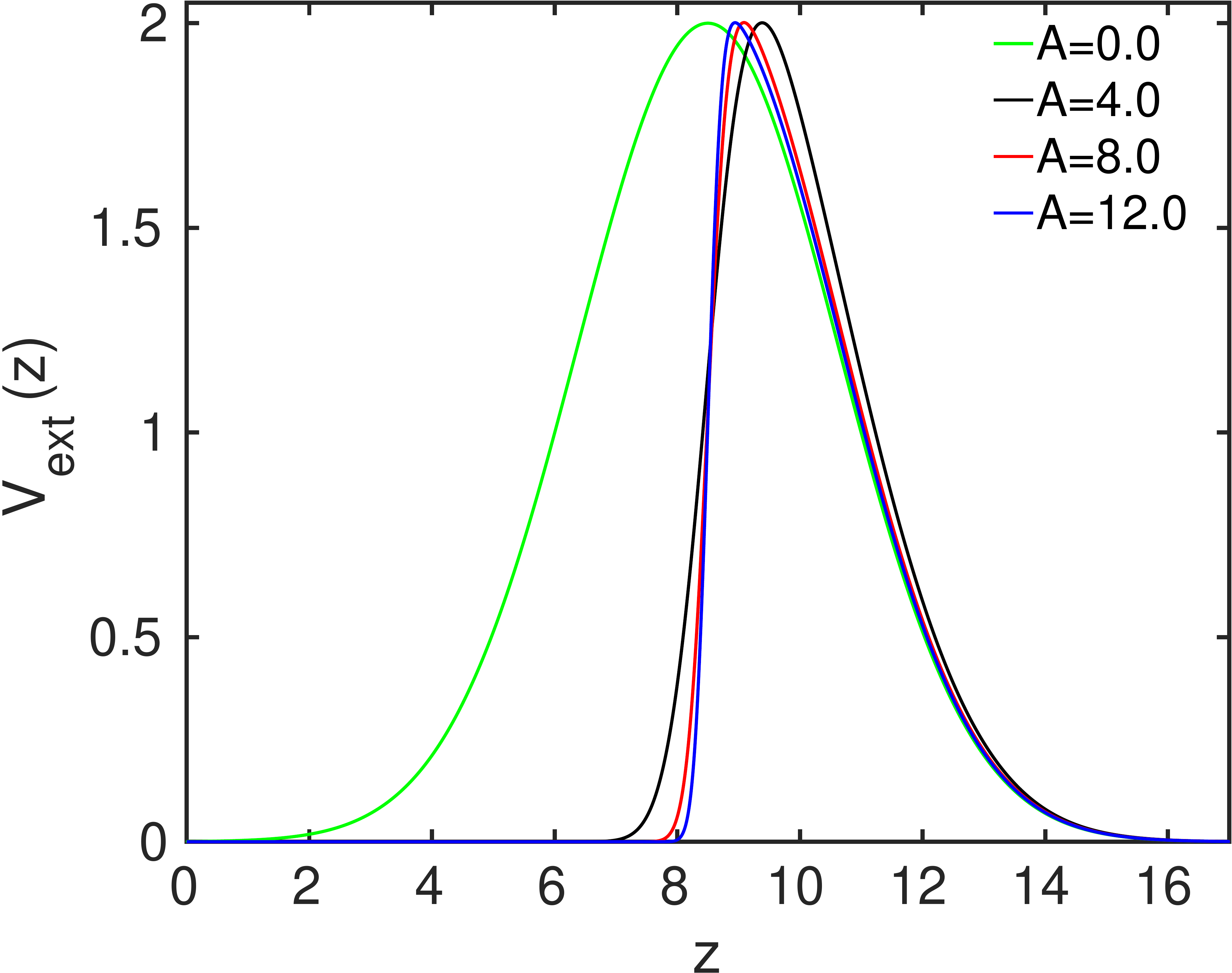}
	\caption{\label{f:asypot} External potential barrier along the $z-$direction
	at $z_0 =L/2$ with varying asymmetry parameter $A=$ 0.0, 4.0, 8.0, and 12.0.
	Note that $A=$ 0.0 is corresponding to the symmetric 
	barrier \cite{j:anil_sym_jcpdyn}; the asymmetry in the potential increases
	with $A$. On increasing the asymmetry, the position of the potential shifts
	little to the right, which does not alter dynamics of the system because
	of the PBCs.}
\end{figure}

We perform constant NVT molecular dynamics simulations of a equi-volume colloidal 
mixture consisting of two different sizes of particles with equal masses. The inter-particle 
interactions in the colloidal mixture are soft and purely repulsive, which are given by
the potential
\begin{equation}
	V(r_{ij}) = \epsilon_{ij}\left(\frac{\sigma_{ij}}{r_{ij}}\right)^{12},
\end{equation}
where $(i,j)\in(s,l)$ are corresponding to the small and large size particles. The
simulation parameters are: $\sigma_{ss}=$ 1.0, $\sigma_{ll}=$ 2.0, $\epsilon_{ss}=$ 1.0, 
and $\epsilon_{ll}=$ 4.0 \cite{j:anil_soft_sym}. Cross interaction parameters are obtained
using Lorentz-Berthelot additive mixture rules, i.e., $\sigma_{sl}=(\sigma_{ss}+\sigma_{ll})/2$
and $\epsilon_{sl}=\sqrt{\epsilon_{ss}\epsilon_{ll}}$. This system is subjected to an external 
asymmetric potential at the center of the box ($z=z_0$) along the $z-$direction, which 
is given as

\begin{equation}
	V(z) = \epsilon_{ext}\,e^{-\left({\frac{z-z_0}{\omega}}\right)^2}\big[1 + 
	\erf\big\{A\left(\frac{z-z_0}{\omega}\right)\big\}\big],
\end{equation}
where $\epsilon_{ext}$, $\omega$, and $A$ are height, width, and asymmetry parameter
of the external potential. We have fixed the width $\omega=$ 3.0, while varying the 
$\epsilon_{ext}$ and $A$ in this study. The asymmetry parameter $A=$ 0 corresponds
to the symmetric external potential \cite{j:anil_soft_sym}, while non-zero values of 
$A$ correspond to the asymmetry in the potential. The simulations are carried out
by varying the asymmetry parameter as $A = $ 0--20, where $\epsilon_{ext}$ is 
adjusted such that the height of the potential remains at 2.0; this potential is 
plotted in Fig. \ref{f:asypot} for few typical values of the asymmetry parameter 
$A$.  Hereafter, we refer the side of asymmetric potential which coincides with
the symmetric Gaussian potential as the \textit{symmetric} side and the one which 
deviates from symmetric Gaussian potential as \textit{asymmetric} side. We simulate
this system at the volume fraction $\phi=$ 0.2 and temperatures $T =$ 2.0--0.3. The
periodic boundary conditions (PBCs) are applied along all the three directions of a
cubic simulation box of length $L=$ 17; detailed simulation information is given 
in \ref{sa:simdet}. To look at the finite-size effects, we performed simulations
of the system at box lengths $L=$ 15, 19, and 21, keeping other simulation 
parameters unchanged, which is given in \ref{sa:finsize}. Dynamical properties
of the system are computed from the phase space trajectories produced in each
production run and averaged over five simulation runs, each starting from a
random initial configuration.

\section{Results and discussion\label{s:results}}
\begin{figure}
	\includegraphics[width=8.5cm, height=10.0cm]{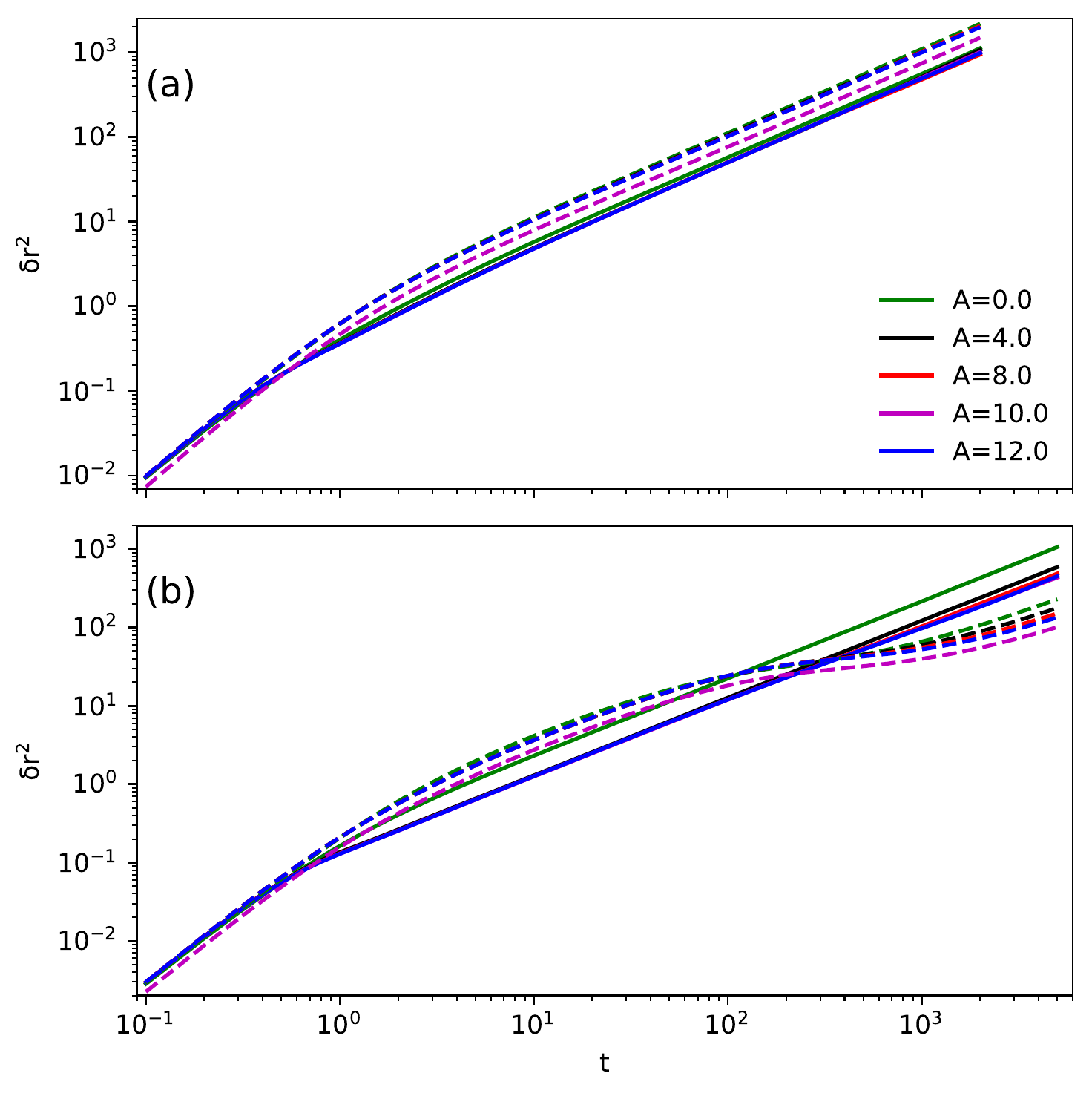}
	\caption{\label{f:msd} Mean-squared displacement of the smaller (dashed lines)
	and the larger (solid lines) particles at $A=$ 0.0, 4.0, 8.0, 10.0, and
	12.0. (a) $T=$ 1.0 and (b) $T=$ 0.3.}
\end{figure}

To examine the effect of asymmetry in the barrier on the dynamics of the colloidal
mixture, we have calculated mean-squared displacement (MSD) of both sizes of
particles along the $z-$direction as
\begin{equation}
	\label{e:msd}
	{\delta r}^2 = \frac{1}{N} \langle (r_z(t) - r_z(0))^2\rangle.
\end{equation}
Since the external potential barrier is only along the $z-$direction and the volume 
fraction is low, the MSD of both sizes of particles does not show jump or caged
like motion at low temperatures along the directions normal to the applied 
external potential barrier for the symmetric \cite{j:anil_sym_jcpdyn} and the 
asymmetric case (see Fig. \ref{f:msdx} of \ref{sa:msdx}). In Fig. \ref{f:msd},
we plot the MSD of both species of particles at $T$ = 1.0 and $T$ = 0.3 at 
different asymmetry parameters. At the higher temperatures, both the species
in the binary mixture undergo normal diffusion at all values of asymmetry 
parameter $A$. As expected, the dynamics of the smaller particles is faster 
than that of the larger particles, as evident from the higher values of MSD 
for the smaller particles. However, at low temperatures, the MSD of the smaller 
particles deviates from the linear behavior and develops a plateau at intermediate 
times. This is due to the localization of the smaller particles between the 
external potential barriers (multiple barriers arise due to the PBCs), as in 
the case of symmetric barriers ($A$ = 0). However, the MSD of the larger 
particles remains linear (at long times) even at the lower temperatures due 
to the attractive depletion interaction between the external potential barrier 
and the larger particles. The asymmetry in the potential barrier does not
affect the qualitative nature of MSD of the smaller particles at all 
temperatures, though it slows down with $A$ that is pronounced at $A=$ 10. 
This slowing down can be dynamical in the sense that the higher asymmetry in
the external potential reduces the probability of larger particles crossing over 
the barrier, thus have to spend a considerably large amount of time near the 
barrier. This, in turn, increases the probability of large particles reversing
the direction of motion near the asymmetric side, therefore, a probability
of finding the smaller particles near the barrier increases. This is evidenced in the 
peak in density profile of the smaller particles near the asymmetric side of 
the barrier [see Figs. \ref{f:dfs}(a--b)]. This interesting observation
will be further investigated. However, the MSD of the larger particles decreases 
with increasing $A$. This is expected as the depletion interaction is between
the potential barrier and the larger particles and any change in the potential
will be mainly affecting the dynamics of the larger particles.

\begin{figure}
	\includegraphics[width=8cm, height=10cm]{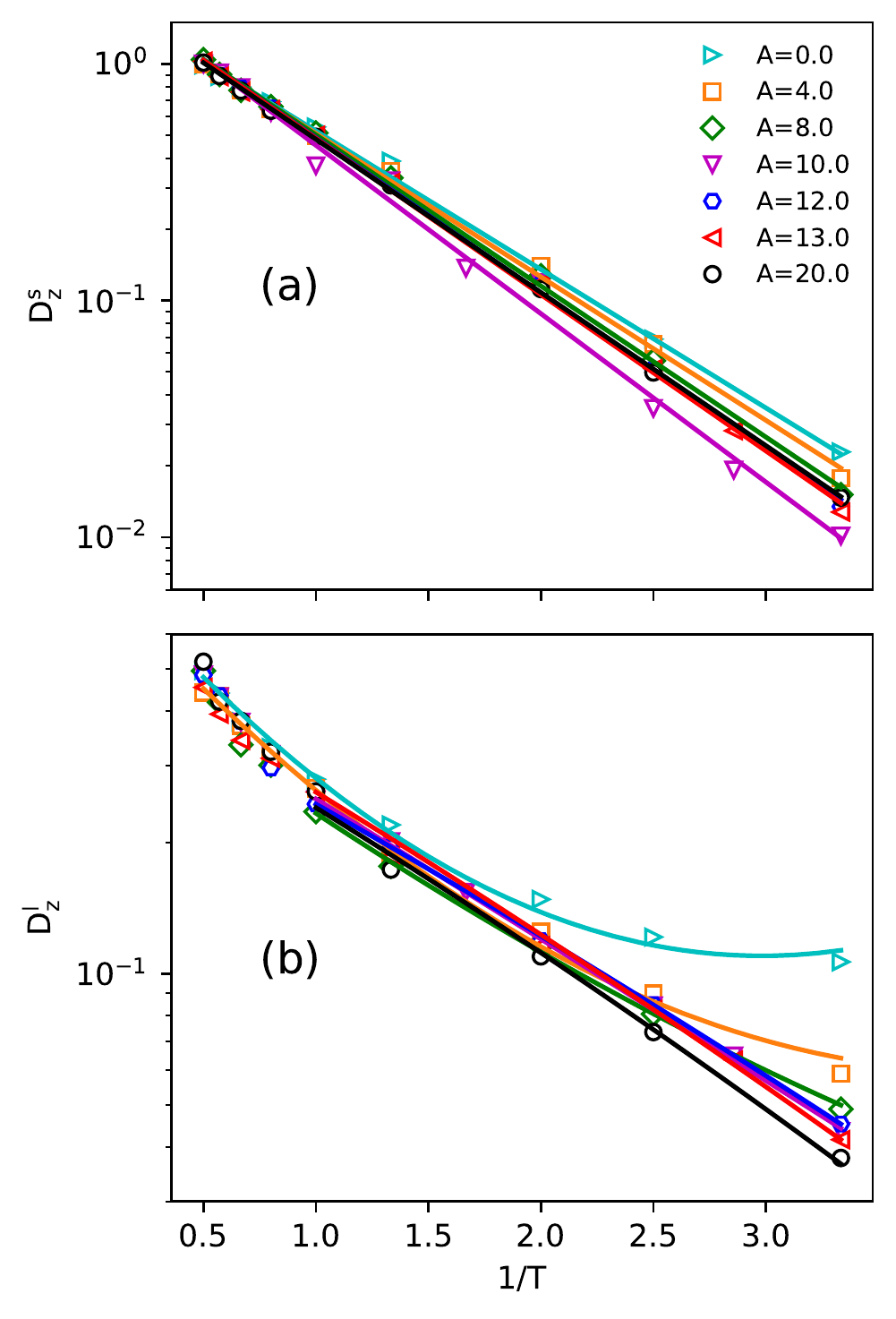}
	\caption{\label{f:dczsl} Diffusion coefficient of the (a) smaller particles
	and the (b) larger particles at temperatures $T=$ 2.0--0.3 along the
	$z-$direction. Solid lines are fit to the data points, whereas symbols 
	correspond to the data at different asymmetry parameters. At $A=$ 8.0 and 
	above, the the diffusion data of the larger particles, is fitted in the 
	temperature range 1.0--0.3.}
\end{figure}

\begin{figure}
	\includegraphics[width=6.5cm, height=5cm]{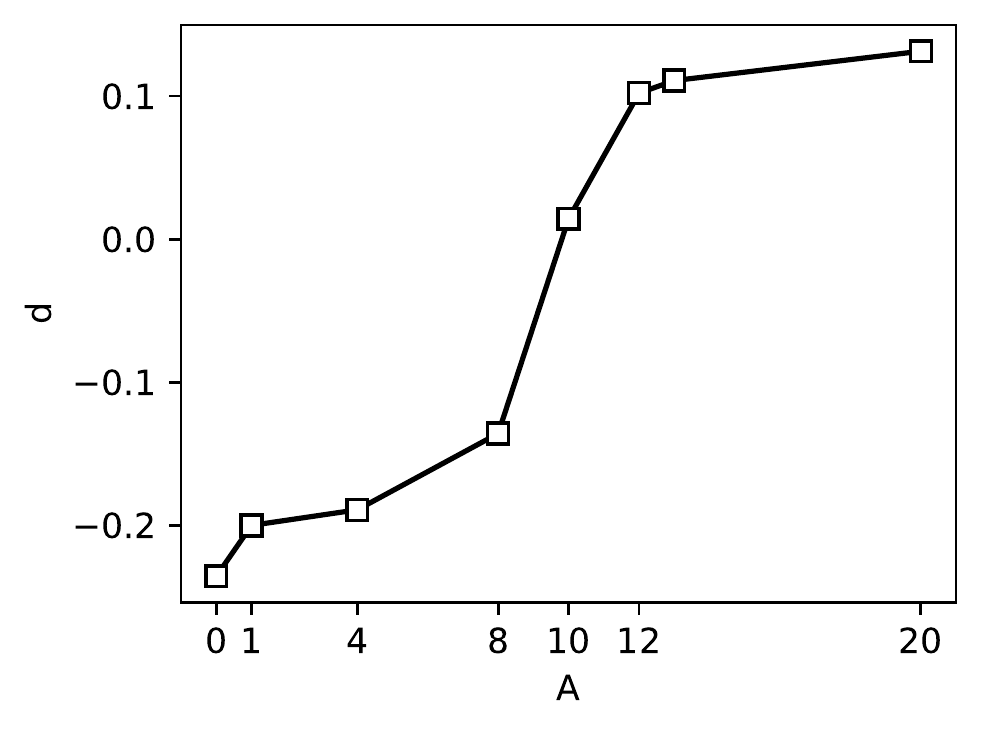}
	\caption{\label{f:deform} Deformation parameter $d$ is obtained from the fitting 
	of the $d-$Arrhenius law (Eq. \ref{e:dArr}) to $\ln(D_z^l)$ \textit{vs} $1/T$
	curves, shown in Fig. \ref{f:dczsl}(b), which is plotted against the asymmetry
	parameter $A$.}
\end{figure}

\begin{figure}
	\includegraphics[width=8cm, height=8.8cm]{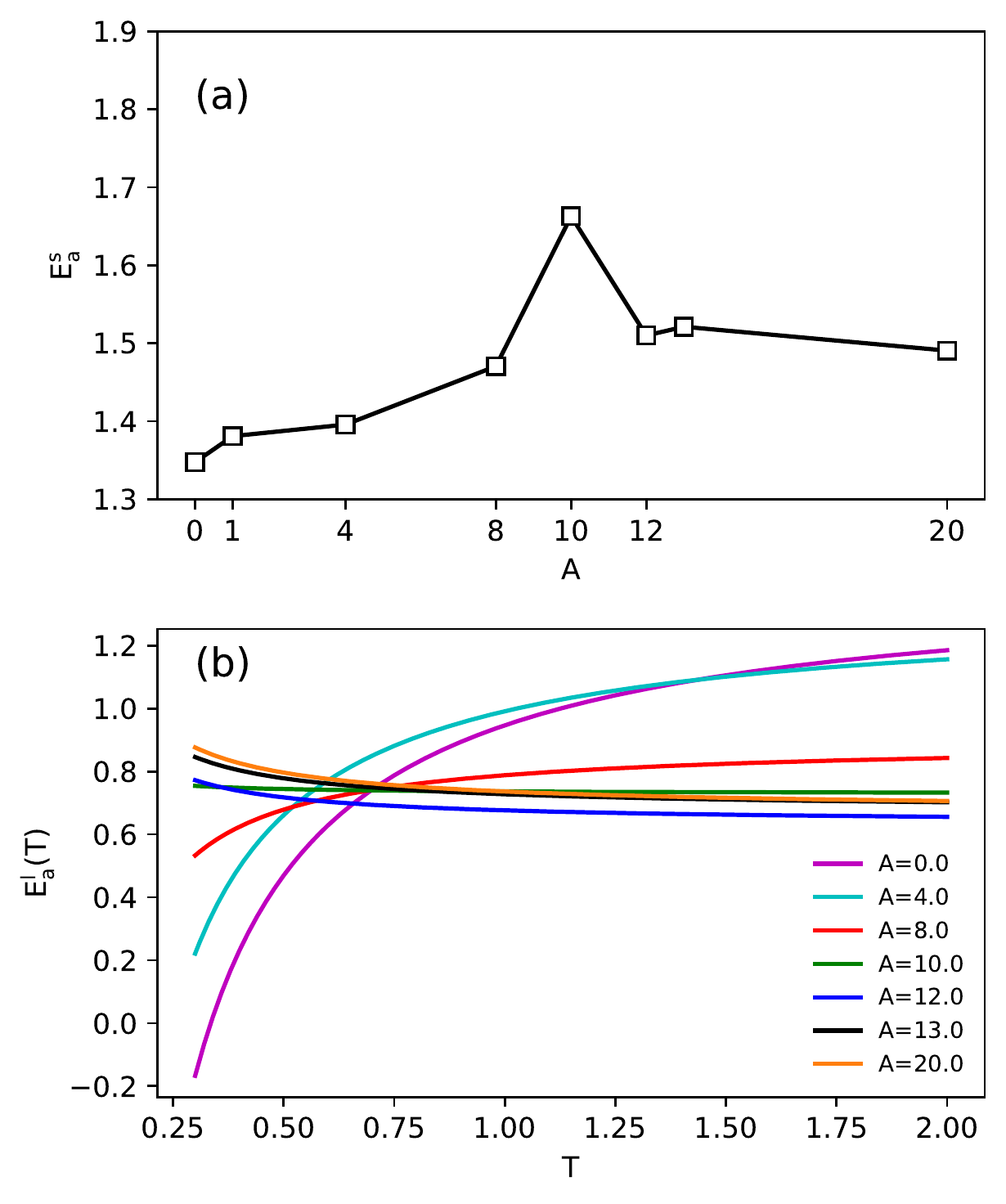}
	\caption{\label{f:acteng} (a) Activation energy of the smaller 
	particles, $E_a^s$, which is temperature independent, is plotted against 
	$A$. It shows a least variation with the asymmetry parameter, except at 
	$A=$ 10. (b) Activation energy of the larger particles $E_a^l(T)$ is
	temperature dependent and showing a crossover (from $A=$ 10 to above) from 
	sub-Arrhenius to super-Arrhenius diffusion.}
\end{figure}

The long time diffusion coefficient of the particles along the direction of external
potential barrier in the colloidal mixture is computed as
$D_z = \lim\limits_{t \to \infty} \frac{{\delta r}^2}{2t}$, where ${\delta r}^2$ is
computed from Eq. \ref{e:msd}. Figure \ref{f:dczsl}(a) shows a semi-log plot 
of $D^s_z$ \textit{vs} the inverse temperature (1/$T$). As temperature
decreases, the diffusion coefficient of the smaller particles decreases rapidly. Here
again the asymmetry in the potential barrier does not affect the diffusivity of the
smaller particles, except for very large values of $A$, namely 10 and 12.
Here again, diffusion of the smaller particles slows down more at $A=$ 10, though 
it is Arrhenius. The plots are linear and can be fitted with the Arrhenius 
equation $D=D_0\exp(-E^s_a/{k_BT})$. This essentially means that activation energy 
for the smaller particles' diffusion is temperature independent. However, the 
diffusion coefficient of the larger particles decreases rather slowly with
temperature, which suggests a non-Arrhenius behavior. As asymmetry in the potential 
barrier increases, $D_z^l$ starts decreasing faster with decreasing temperature, but
the numerical values are still larger compared to $D_z^s$.
The temperature dependence of $D_z^l$ is found to be sub-Arrhenius in the case of 
symmetric potential barrier \cite{j:anil_soft_sym}. In fact, a general $d-$Arrhenius
equation has been proposed to include the deviations from Arrhenius behavior in
diffusion, especially at lower temperatures, as
\begin{equation}
	\label{e:dArr}
	D(T) = A\Big[1-d\frac{E_0}{k_BT}\Big]^{1/d},
\end{equation}
\noindent where $E_0$ is the height of the barrier and $d$ is the deformation 
parameter \cite{j:aquilanti}. For $d$ = 0, above equation tends to the original
Arrhenius equation. For positive values of $d$, $D \, \, vs \, \, 1/T$ curve will
be convex and the behavior is known as the super-Arrhenius diffusion. The most common
mechanism known for the super-Arrhenius diffusion is the correlated dynamics of particles
such as in the supercooled liquids \cite{wales1,wales2}. For negative values of $d$, the 
$D \, \, vs \, \, 1/T$ curve will be concave and the diffusion is said to be sub-Arrhenius.
As pointed out in the introduction, the sub-Arrhenius behavior is mostly observed in 
quantum systems, where quantum tunneling plays a crucial role in the process. From 
Fig. \ref{f:dczsl}(b), it is clear that the dynamics of the larger particles is 
sub-Arrhenius at low values of asymmetry in the potential barrier. However, as the 
asymmetry increases the concave nature of the curve diminishes and become linear or 
convex. To find this, we have fitted the $d-$Arrhenius equation and obtained the $d$ 
values at different values of $A$, which is shown in Fig. \ref{f:deform}.
At low values of $A$, $d$ is negative, a characteristic of the sub-Arrhenius diffusion.
However, as $A$ increases, the deformation parameter $d$ increases and for the larger
values of $A$ it crosses over to positive values (though small), indicating a crossover
from sub-Arrhenius to super-Arrhenius diffusion. On further increasing the 
asymmetry in the external potential, the $d$ value reached a plateau. This is because
the shape of the external potential does not change significantly, at the higher 
$A$ (see Fig. \ref{f:asypot}), and the volume fraction of the system is very low.
We expect that super-Arrhenius diffusion for the larger particles of the 
mixture, subjected to the asymmetric external potential barrier, may enhance at 
the higher system volume fractions, well below $\phi_g$.
We have calculated the activation energy from the diffusion coefficient of both 
species of particles by fitting the Arrhenius (for the smaller particles) 
or $d-$Arrhenius (for the larger particles) equations from $D \, \, vs \, \, 1/T$ 
curve; these are plotted in Figs. \ref{f:acteng}(a--b). The activation energy of
the smaller particles shows a little increment with $A$. Since the diffusion of the
larger particles is $d-$Arrhenius, the activation energy for diffusion is
temperature dependent and hence plotted against temperature for different values 
of $A$. As temperature decreases the activation energy decreases for the smaller 
values of $A$, however the trend gets reversed for the larger particles at the 
higher values of $A$. This is consistent with the change in the sign of $d$, again
suggesting the crossover from sub-Arrhenius to super-Arrhenius diffusion.
\begin{figure}
	\includegraphics[width=8cm, height=9cm]{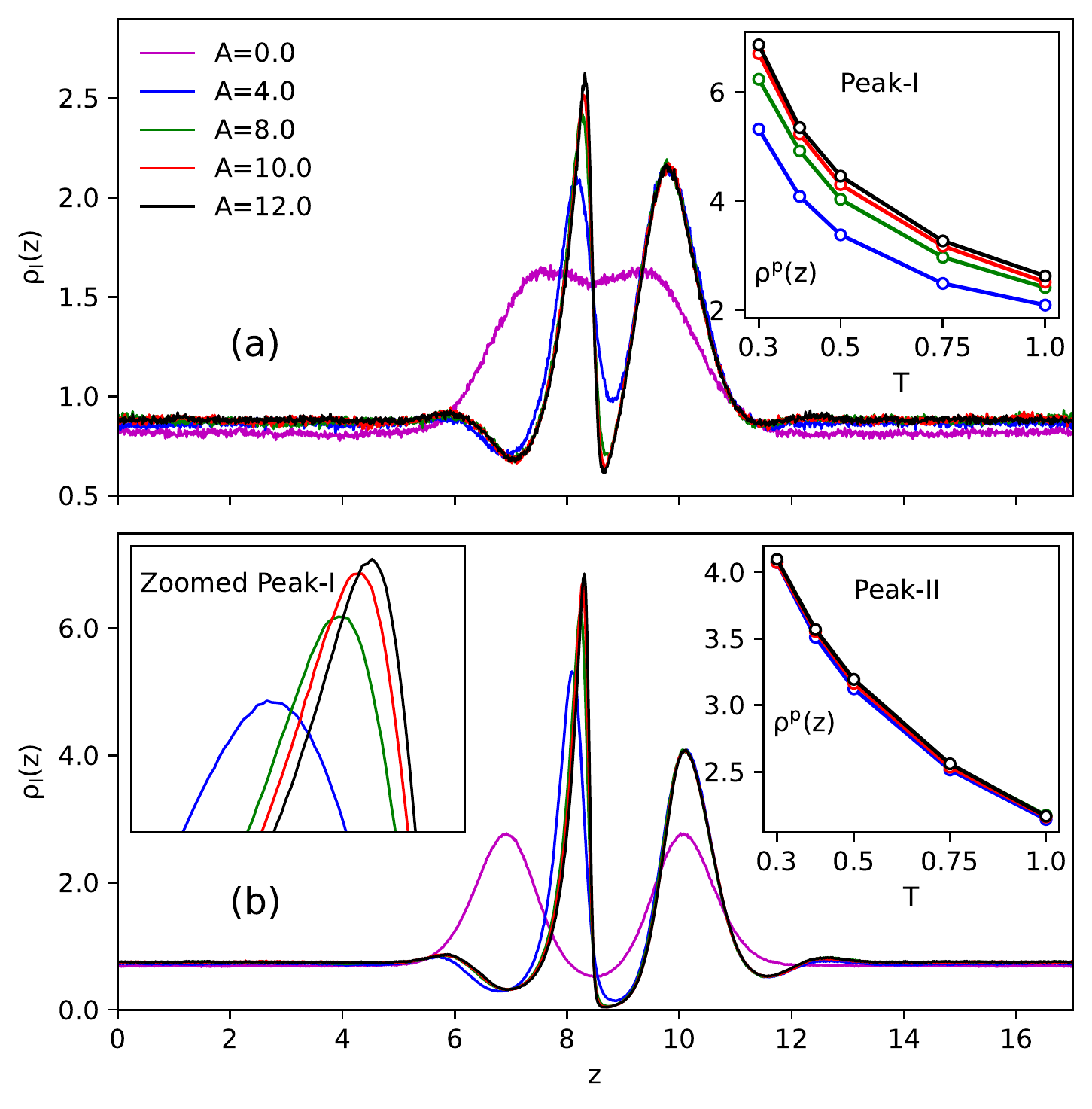}
	\caption{\label{f:dfl} Density profile of the larger particles along the
	$z-$direction at $A=$ 0.0, 4.0, 8.0, 10.0, and 12.0. 
	(a) $T=$ 1.0 and (b) $T=$ 0.3. Top right insets of (a) and (b) show
	plots of peak heights of $\rho_l(z)$ \textit{vs} temperature $T$ at 
	$A=$ 4.0, 8.0, 10.0, and 12.0; these peaks are named as 
	peak I (asymmetric side) and peak II (symmetric side). The zoomed area
	of the peak I at $T=$ 0.3 is shown in the top left inset of (b), showing 
	an increment in its height with position shifting towards right. This
	signifies an accumulation of the larger particles towards the asymmetric 
	side of the barrier.}
\end{figure}
\begin{figure}
	\includegraphics[width=8cm, height=11cm]{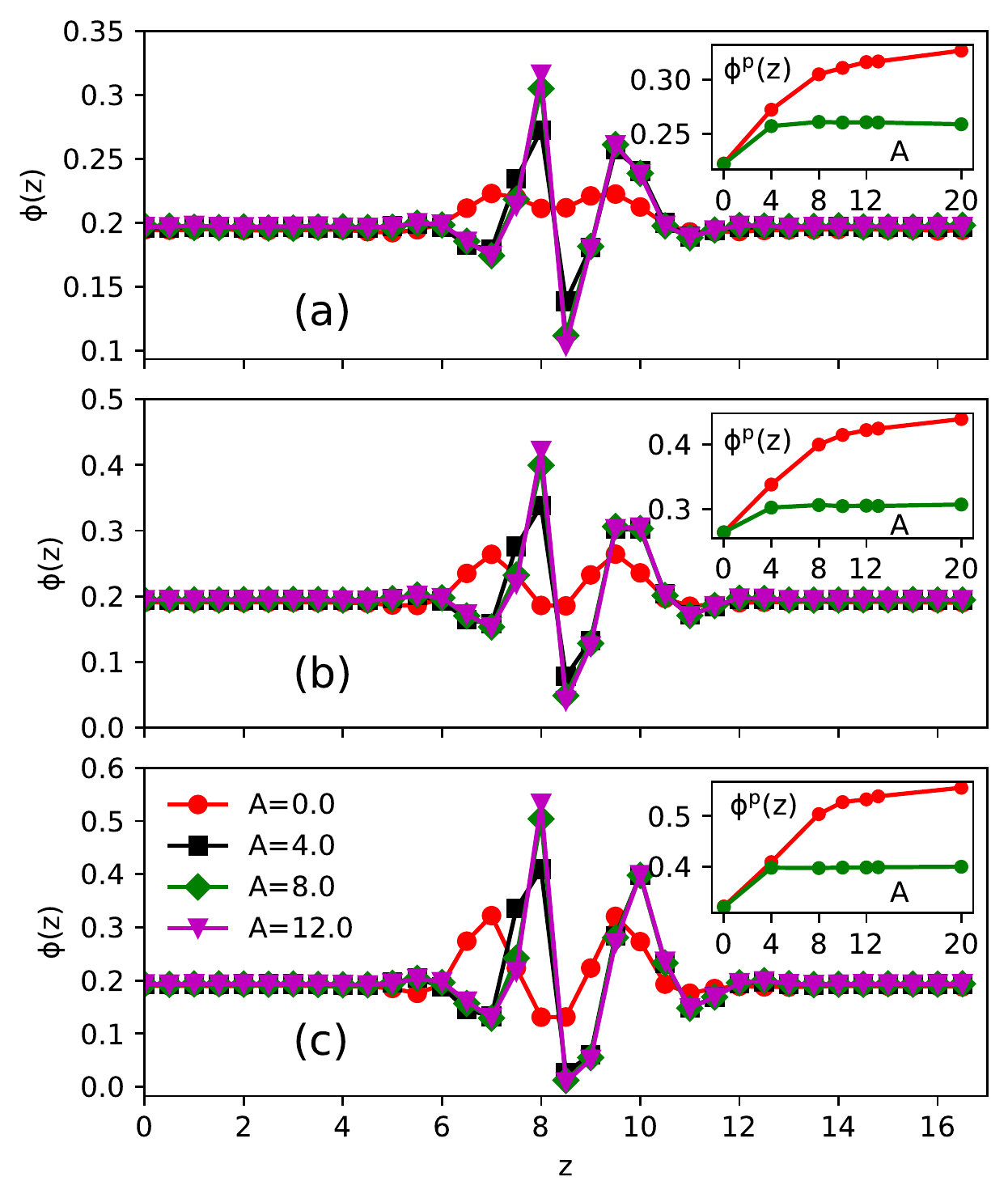}
	\caption{\label{f:phiz} Effective volume fraction of the system (over all 
	particles) along $z-$direction at asymmetry parameters $A=$ 0.0, 4.0, 8.0, 12.0. 
	(a) $T=$ 1.0, (b) $T=$ 0.5, and (c) $T=$ 0.3. Legend of (c) is applied 
	to (a) and (b). Insets of (a--c) show a variation in peak heights of $\phi(z)$, 
	i.e., $\phi^p(z)$, on the asymmetric (\textit{red}) and 
	symmetric (\textit{green}) side of the barrier, against the asymmetry parameter 
	$A$. The increment and constant values of $\phi^p(z)$, at and above $A=$ 4.0,
	on the asymmetric and symmetric side of the barrier, respectively, corroborate
	with the enhancement of the peaks of $\rho_l(z)$.}
\end{figure}
\begin{figure}
	\includegraphics[width=7.3cm, height=10cm]{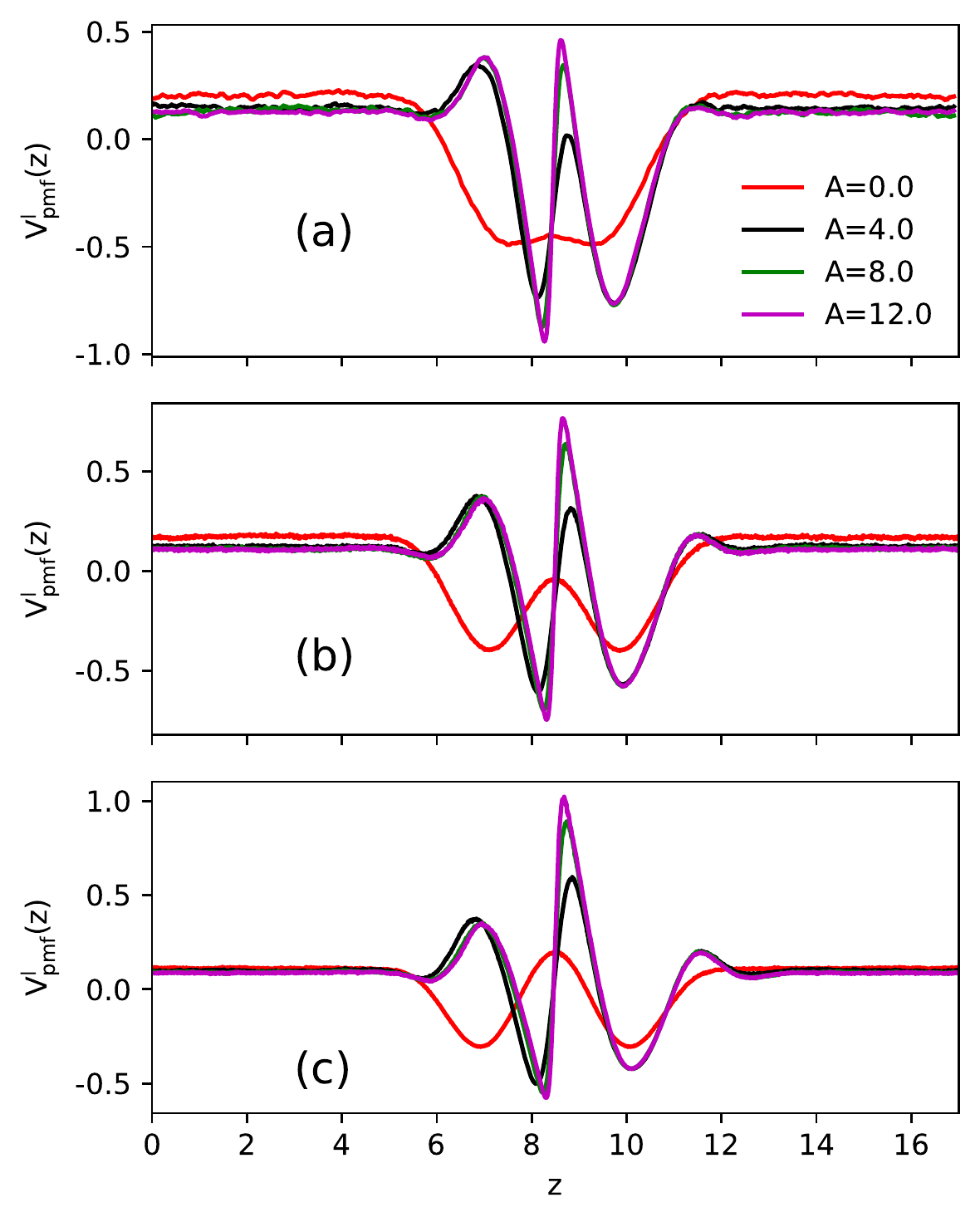}
	\caption{\label{f:pmf} Potential of mean force of the larger particles 
	along $z-$direction at asymmetry parameters $A=$ 0.0, 4.0, 8.0, 12.0. 
	(a) $T=$ 1.0, (b) $T=$ 0.5, and (c) $T=$ 0.3. Legend of (a) is applied 
	to (b) and (c).}
\end{figure}

To gain further insight into this effect of asymmetry on the diffusive behavior
of larger particles, we calculate their density profile, $\rho_l(z)$, along the 
$z-$direction. This is plotted in Figs. \ref{f:dfl}(a--b) for two representative
temperatures $T$ = 1.0 and $T$ = 0.3. The density profile shows peaks in 
the neighborhood of the external potential barrier on either side. This is due to 
the attractive depletion interaction between the barrier and the larger particles.
For the symmetric barrier ($A$ = 0), the peaks are symmetric around the center 
of the potential barrier, as expected. However, as $A$ increases, the density 
profile also becomes asymmetric. Moreover, local density increases at the 
asymmetric side of the potential compared to the symmetric side. For more 
clarity, this is shown in the left inset of Fig. \ref{f:dfl}(b). As temperature
is lowered, the peak height increases for both peaks of $\rho_l(z)$ at all $A$. 
However, the local density at the centre of the potential barrier decreases and 
approaches close to zero at very low temperatures. To estimate the effective 
density of particles near and off the barrier, we calculate effective volume 
fraction along $z-$direction as
$\phi(z) = \frac{\pi}{6L^2dz}[N_s(z)\sigma_{ss}^3 + N_l(z)\sigma_{ll}^3]$ 
by slicing the three dimensional simulation box at width $dz=$ 0.5 and
calculating number of small ($N_s$) and 
large ($N_l$) particles along $z-$direction; this is shown in Figs. \ref{f:phiz}(a--c).
The $\phi(z)\simeq \phi$ (= 0.2) at all $A$ away from the barrier, shown at three 
representative temperatures. However, $\phi(z)$ differs from $\phi$ near the barrier.
It increases equally on both sides of the barrier for $A=$ 0 ($\phi^p(z)$ are identical
in the insets of Figs. \ref{f:phiz}(a--c)), which becomes asymmetric with an 
increasing asymmetry in the potential. This is clearly shown by plotting
$\phi^p(z)$ \textit{vs} $A$, where the peak height at the symmetric side of the 
barrier becomes constant from $A=$ 4 onwards. On the other hand, the $\phi^p(z)$ 
at the asymmetric side of the barrier grows faster with $A$ up to $A=$ 8, its growth
slows down beyond it. Interestingly, at $T=$ 0.3, $\phi(z)$ exceeds 
0.5 (or $\frac{\phi(z)}{\phi} >$ 2.5) beyond $A=$ 8.0, approaches $\phi(z)\simeq$ 
0.56 (at $A=$ 20) that is near the glass volume fraction for hard spheres,
i.e., $\phi_g=$ 0.58 \cite{megen_HSGlass}. The crossover from sub-Arrhenius to
super-Arrhenius diffusion is also observed beyond $A=$ 8 (see Fig. \ref{f:deform}). 
Studies on the role of system density on the kinetic fragility and the configurational
entropy show that the kinetic fragility increases whereas the configurational entropy
decreases with density, showing a rapid slowing down at the higher densities, one of 
the key signatures of the supercooled liquids \cite{sastry-ds,sastry-ds1}. An 
accumulation of the particles towards the asymmetric side of the barrier above 
$A=$ 8.0 suggests a frequency of crossing over the barrier decreases with
temperature, thus increases local density in the proximity of the barrier. We 
calculate potential of mean force for the larger 
particles, $V_{pmf}^l(z)=-k_B \, T\ln[\rho_l(z)]$, to examine this depletion 
effect due to the barrier. Figures \ref{f:pmf}(a-c) compare $V_{pmf}(z)$ of the larger 
particles at the asymmetry parameters $A=$ 0.0, 4.0, 8.0, 12.0 at one higher ($T=$ 1.0) 
and two lower ($T=$ 0.5, 0.3) temperatures; $V_{pmf}^l(z)$ varies much with $T$ and $A$.
When the barrier is symmetric, the larger particles are crossing it from both the sides 
with equal probabilities, even at low temperatures because the height of $V_{pmf}^l(z)$
is nearly zero (more details can be found in 
Refs. \cite{j:anil_sym_jcpdyn,j:anil_soft_sym}). As $A$ increases, the height as well
as the asymmetry in the potential of mean force increases, and the larger particles 
start accumulating (moreover on the asymmetric side) near the barrier. Since crossing
the barrier becomes increasingly difficult for the larger particles at lower 
temperatures, they aggregate near the barrier on both sides (more on the asymmetric 
side), which leads to the higher local density of the larger particles. However, 
there are no signatures of local freezing of layers and crystallization of the 
particles near the barrier, which is examined by two-dimensional structure factor 
and shown in Fig. \ref{f:skz} of \ref{sa:sk}, to keep the continuation of our main
results. The smaller particles remain (mostly) away from the barrier, which is due
to the depletion effect in the binary mixture caused by the symmetric external 
potential barrier \cite{j:anil_sym_jcpdyn,j:anil_soft_sym}. This in turn leads to 
local caging and slowing down of the dynamics of the larger particles, which 
is examined by calculating the non-Gaussian parameter (NGP), the incoherent 
intermediate scattering function, and the self part of the van-Hove correlation
function.
\begin{figure}
	\includegraphics[width=7.5cm, height=6.0cm]{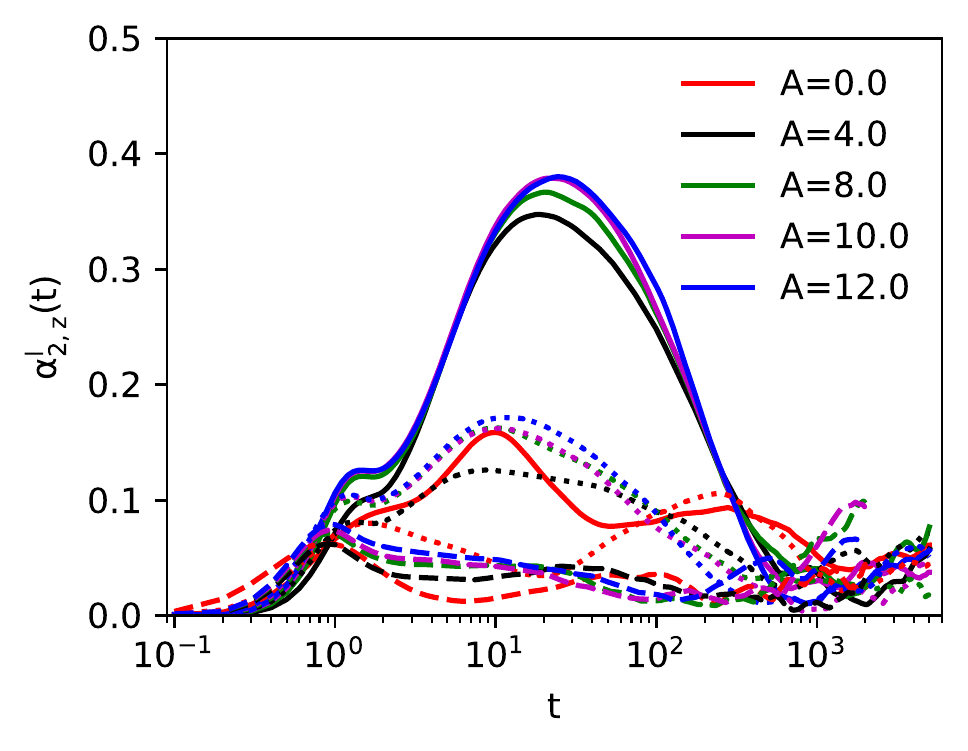}
	\caption{\label{f:ngpl} Non-Gaussian parameter of the larger particles along the 
	applied external potential barrier at $A=$ 0.0, 4.0, 8.0, 10.0, 12.0, and 
	temperatures $T=$ 1.0 (dashed lines), 0.5 (dotted lines), and 0.3 (solid lines).}
\end{figure}

In the case of the symmetric potential barrier, the NGP of the larger particles
is (approximately) zero, which indicates that these particles are not localized
between the barriers and their dynamics is diffusive even at low
temperatures \cite{j:anil_sym_jcpdyn}. In contrast to this, the NGP of the
smaller particles shows marked deviations from zero at intermediate times as we 
decrease the temperature. This corresponds to the localization of the smaller 
particles between the potential barriers resulting in a plateau in MSD at 
intermediate times \cite{j:anil_sym_jcpdyn}. From the MSD's of the smaller 
particles [see Fig. \ref{f:msd}(a)], it is evident that their dynamics does not 
change qualitatively, though changes quantitatively, with $A$.  Therefore, we
show the NGP of the larger particles only (see Fig. \ref{f:ngpl}) to examine 
the cagelike features due to their localization near the barrier, which is 
defined as
\begin{equation}
	\alpha^l_{2,z}(t) = \frac{1}{3}\frac{\langle (r_z(t) -
	r_z(0))^4\rangle}{{\langle(r_z(t)- r_z(0))^2\rangle}^2} -1.0  .
\end{equation}
Figure \ref{f:ngpl} shows that NGP of the larger 
particles is below the value 0.1 at temperature $T=$ 1.0, whereas it starts growing
from $T=$ 0.5 and $A=$ 4.0 onwards. Interestingly, from $A=$ 4.0 of $T=$ 0.5, 
$\alpha^l_{2,z}(t)$ shows two peaks --- first one at time $t\approx$ 1.0, while second 
one at time $t\approx$ 10.0 that is shifted towards the longer times at low temperatures 
and the larger $A$. At $T=$ 0.3, both peaks of NGP grow, which
is a clear signature of the onset of the cagelike motion found in supercooled liquids; 
the time scales of both peaks are akin to the $\beta$ and $\alpha-$relaxation
time scales in the dynamics of supercooled liquids \cite{b:berthierdh,j:karmakar_beta}.
Thus, these peaks show the cagelike motion of the larger particles near the barrier,
which enhances with the asymmetry parameter $A$, therefore, the larger particles exhibit
super-Arrhenius diffusion at $A=$ 10.0 onwards. On contrary, the large particles' NGP does 
not show any growth up to $A=1.0$, where sub-Arrhenius diffusion is observed.
\begin{figure}
	\includegraphics[width=8cm, height=9cm]{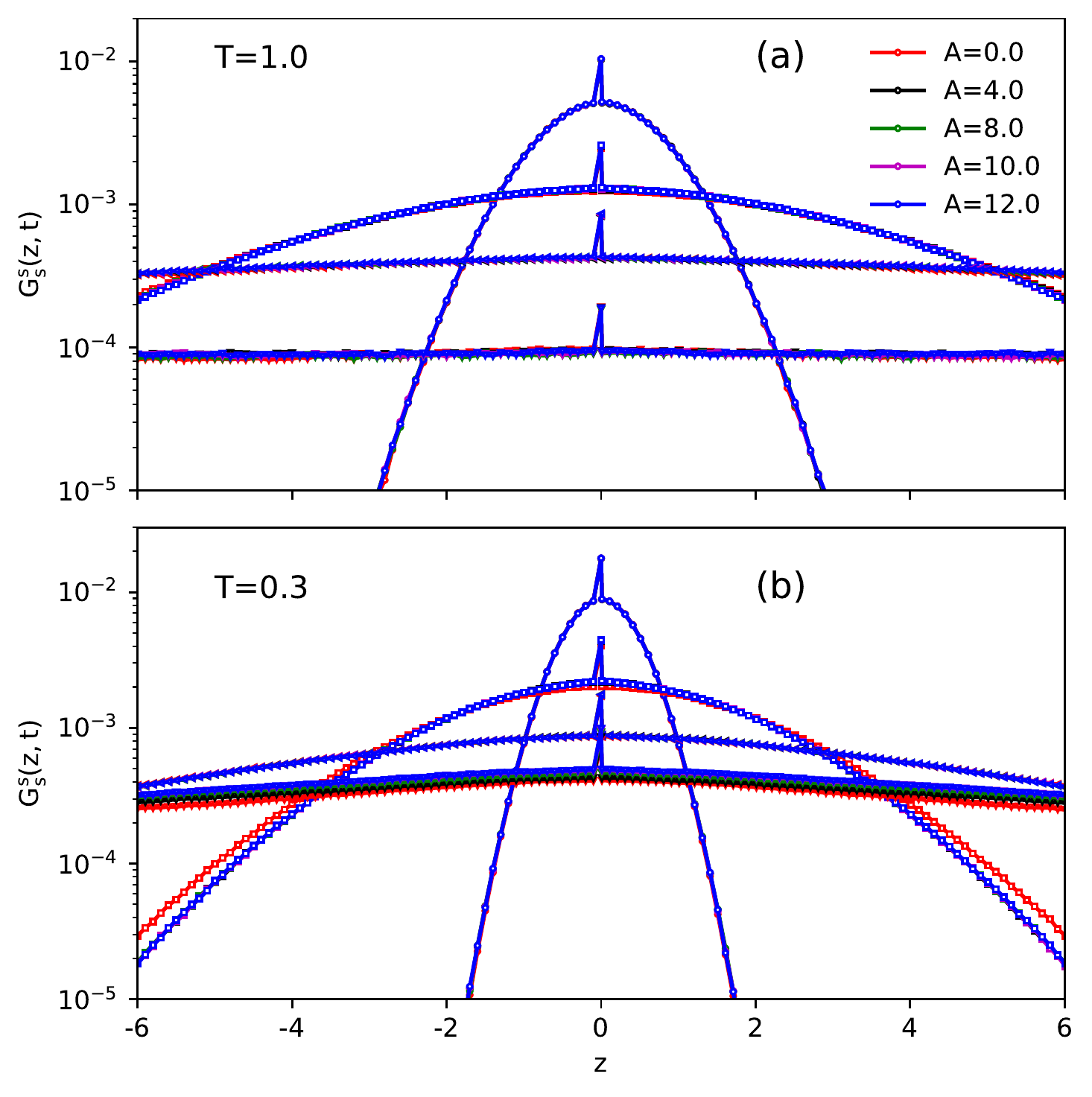}
	\caption{\label{f:gszts} Self part of van-Hove correlation function of the smaller
	particles at $A=$ 0.0, 4.0, 8.0, 10.0, 12.0, and temperatures (a) $T=$ 1.0 and
	(b) $T=$ 0.3. The symbols $\bigcirc$, $\square$, $\vartriangleleft$, and 
	$\bigtriangledown$ ($T=$ 0.3) correspond to times $t=$ 1.0, 10.0, 100.0, and 
	2000.0 (5000.0).}
\end{figure}
\begin{figure}
	\includegraphics[width=8.5cm, height=9.5cm]{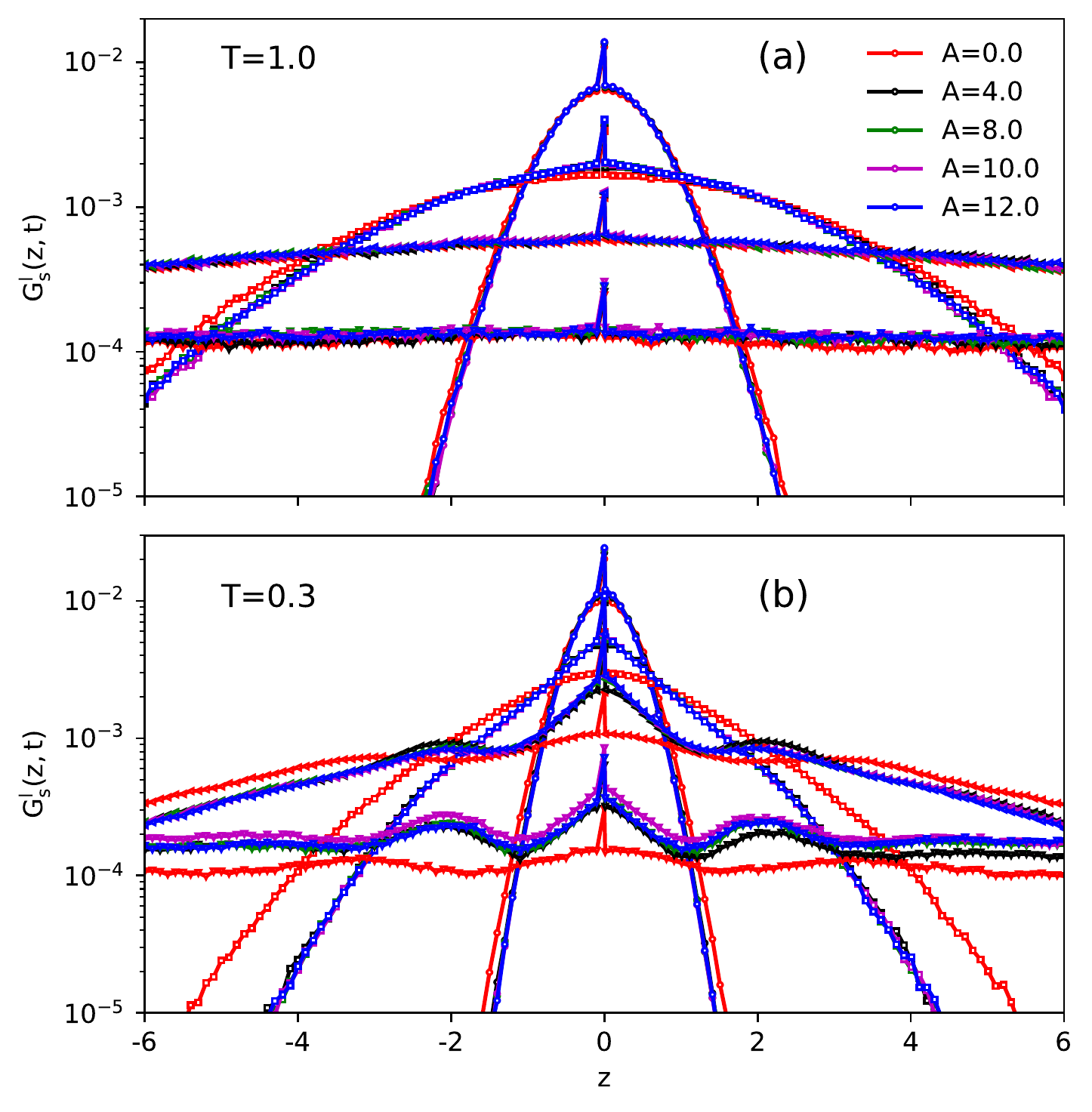}
	\caption{\label{f:gsztl} Self part of van-Hove correlation function of the larger
	particles at $A=$ 0.0, 1.0, 4.0, 8.0, 10.0, and temperatures (a) $T=$ 1.0 and
	(b) $T=$ 0.3. The symbols $\bigcirc$, $\square$, $\vartriangleleft$, and 
	$\bigtriangledown$ ($T=$ 0.3) correspond to times $t=$ 1.0, 10.0, 100.0, and 
	2000.0 (5000.0).}
\end{figure}
\begin{figure*}
	\centering
	\includegraphics[width=15.5cm, height=10.5cm]{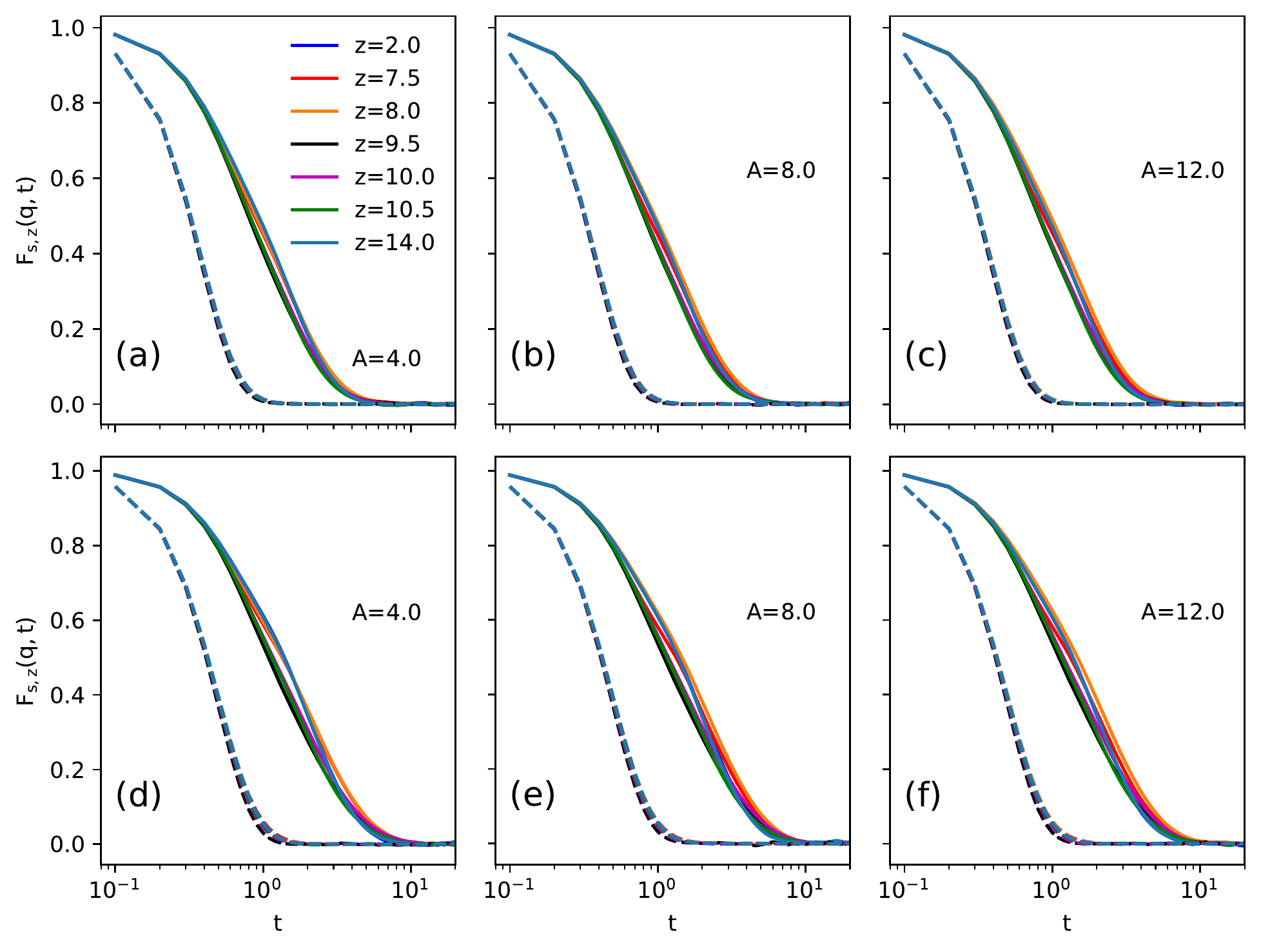}
	\caption{\label{f:fszkt} Incoherent intermediate scattering function of the
	smaller (dashed lines) and the larger (solid lines) particles at low 
	temperatures, $T=$ 0.5 (a--c) and $T=$ 0.3 (d--f), near and off the barrier,
	at varying asymmetry. This shows that dynamics of the larger particles is
	slower than the smaller particles near the barrier. Interestingly, a little
	hump appears in $F_{s,z}(q,t)$ near the asymmetric side of the barrier, 
	showing their cage like motion.}
\end{figure*}
\begin{figure}
	\includegraphics[width=8.5cm, height=7.0cm]{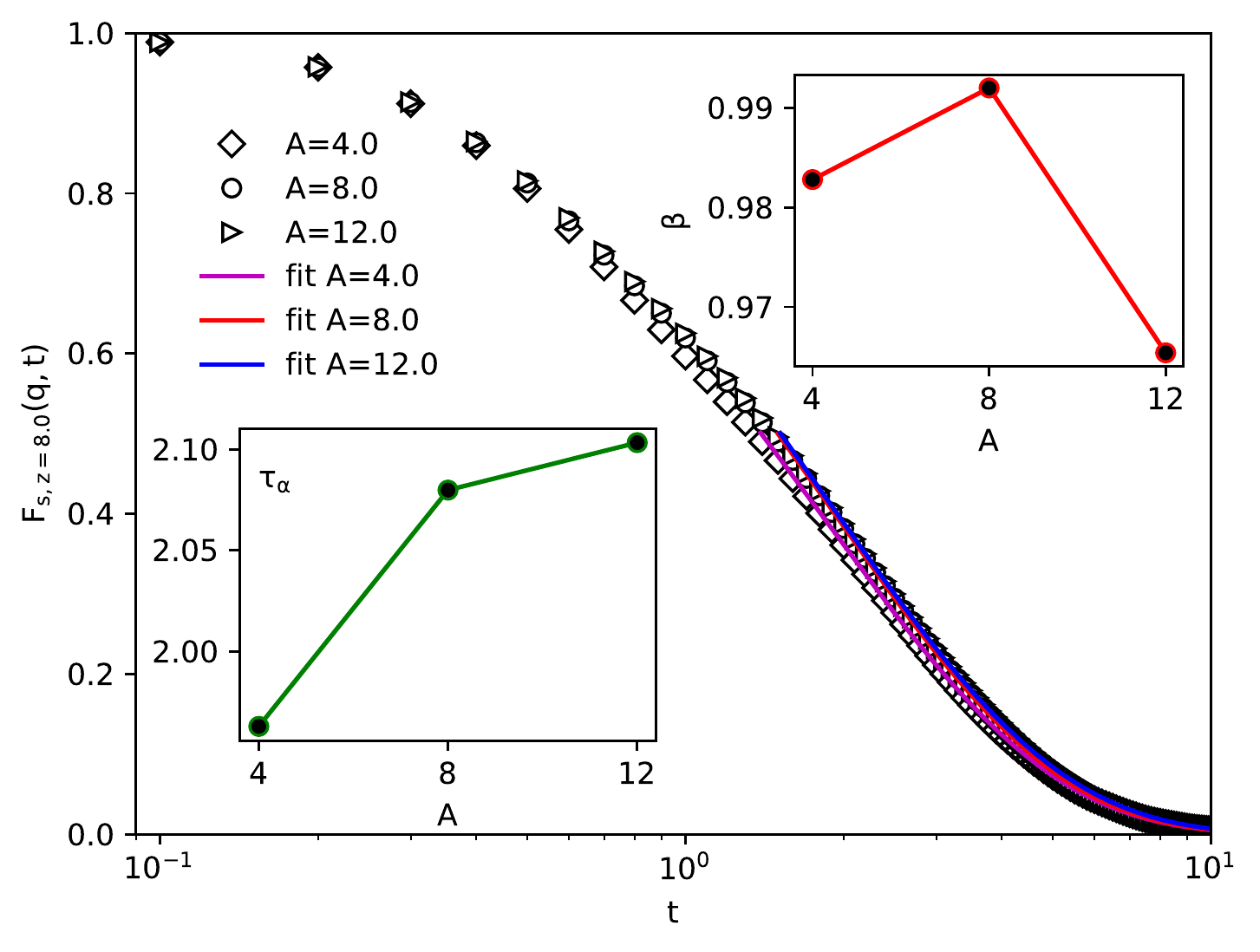}
	\caption{\label{f:fszktl} Self intermediate scattering function 
	is plotted against time $t$ at $z=$ 8.0 and $T=$ 0.3, which is 
	fitted with the empirical Kohlrausch-Williams-Watts (KWW) function 
	of the form $f(t) \propto \exp[-(t/\tau)^\beta]$
	\cite{kwwNgai,berthierRMP,polyJS}. The exponent $\beta$ and the 
	relaxation time $\tau_\alpha$, at $z=$ 8.0, are plotted in the top-right and 
	bottom-left insets, respectively.}
\end{figure}

The MSD of the larger particles shows sub-Arrhenius diffusion along the
$z-$direction due to crossing of the barrier comprising large jumps over it, which
decreases with temperature and asymmetry parameter $A$. At low temperatures and
the larger $A$, the large particles become caged due to their localization near 
the external barrier, preferably at the asymmetric side. The cage and the jump
like motions together can be examined from the self part of the van-Hove correlation 
function along $z-$direction \cite{ham}, which is defined as
\begin{equation}
	\label{e:gszt}
	G_s(z,t) = \frac{1}{N} \left\langle \sum_{i=1}^N \delta[z-z_i(t)
	+ z_i(0)] \right\rangle .
\end{equation}
A plot of $G_s(z,t)$ of small particles at $T=$ 1.0 and 0.3 is shown in 
Figs. \ref{f:gszts}(a--b), showing that $G_s^s(z,t)$ does not depend on the 
asymmetry parameter at all times; at long times tail of $G_s^s(z,t)$ spreads 
because the squared displacement increases with time. At low temperatures,
e.g., $T=$ 0.3, $G_s^s(z,t)$ is independent of the asymmetry parameter $A$,
except at $t=$ 5000.0, where it shows a marginal dependency on $A$. This agrees
with the discussion of MSD of the smaller particles, where its trend does not 
change with $A$, except at long times where marginal variations are shown.
Similar to the smaller particles, $G_s(z,t)$ of the larger particles also
does not depend on the asymmetry parameter $A$ at high 
temperatures (one representative temperature, i.e., $T=$ 1.0 is shown) and the 
shorter time scales, though it changes marginally at long time 
scales [see Fig. \ref{f:gsztl}(a)]. This is supported from the MSD of the 
larger particles, and their density profile that does not change much on 
increasing $A$ at high temperatures --- as the peak heights increase
marginally [see inset of Figs. \ref{f:dfl}(a--b)]. However, at low temperatures,
$G_s^l(z,t)$ varies with the asymmetry parameter $A$ from $t\approx$ 10.0, 
significantly. Interestingly, $G_s^l(z,t)$ start showing a peak from
$t\approx$ 100.0, which start appearing at $z\simeq2.0\sigma_{ll}$ for the
symmetric potential barrier ($A=$ 0), which entails that the larger particles
can hop large distances of the order of twice of their diameter. This shows 
that the larger particles show MSD larger than the expected 
from $\langle(r_z(t)- r_z(0))^2\rangle = 2\,D_z\,t$ , which causes their
diffusion to be sub-Arrhenius. The potential of mean force $V_{pmf}(z)$ shows 
a small effective barrier height ($\epsilon_{ext}^{eff}\simeq$ 0.2) at low
temperatures of $A=$ 0, which means that larger particles can jump over the 
barrier with a small energy penalty. This peak of $G_s^l(z,t)$ shifts toward the 
smaller distances with the increment in the asymmetry of the external potential.
Interestingly, the peak of $G_s^l(z,t)$ starts approaching toward 
$z\simeq \sigma_{ll}$ from $A=$ 4.0 onwards that reaches at $z\simeq \sigma_{ll}$
for $A=$ 10.0, indicating the jumplike motion of the larger particles from the
transient cages, similar to one of the characteristics of supercooled liquids.
The effective potential barrier height for the larger particles at $T=$ 0.3 and
$A=$ 10.0 is $\epsilon_{ext}^{eff}\simeq$ 1.0, which is five times greater than
the height at $T=$ 0.3 and $A=$ 0.0 (symmetric potential). This is a clear 
signature of the localization at larger $A$ and low temperatures due to the
asymmetric external potential barrier, which alters the diffusion of the larger 
particles over the barrier from sub-Arrhenius to super-Arrhenius. Interestingly, 
the oscillating peaks of $G_s^l(z,t)$ are almost identical for positive and 
negative displacements along the applied potential barrier, indicating the 
bidirectional transport of the larger particles across it. 

Many studies where particles accumulate near a confining wall, show slow 
relaxation dynamics due to crowding \cite{wallSlow,tanakaWall,reichmanWall}.
These crowded particles are caged by the surrounding particles, exhibiting 
slow structural relaxation. It is well established that the 
self-intermediate scattering function estimates the relaxation dynamics
of caged particles. Therefore, the accumulation of particles, in the 
vicinity of the barrier, is also examined from the (z-dependent) 
incoherent intermediate scattering function \cite{wallSlow}, which is
calculated as 
\begin{equation}
	\label{e:fszkt}
	F_{s,z}(q,t) = \frac{1}{N} \left\langle \sum_{i=1}^N
	e^{-i {\mathbf q}.\left[{\mathbf r}_i(t) - {\mathbf r}_i(0)\right]} 
	\, \delta[z - z_i(0)] \right\rangle .
\end{equation}
Here, wave numbers $q=$ 5.4 and 2.8 are corresponding to peaks of the static
structure factor for the smaller and the larger particles, respectively.
The $z_i(0)$ is the z-coordinate of an $i$th particle at time $t=$ 0.
Figs. \ref{f:fszkt}(a--c) shows that the relaxation dynamics of the smaller 
particles is almost identical in the sub-Arrhenius ($A < $ 10) and the 
super-Arrhenius ($A >$ 8) regimes of the system, near and far from the 
barrier, even at the lowest temperature ($T=$ 0.3) of the study. This is 
consistent with their mean-squared displacement, diffusion coefficient, 
van Hove correlation function, and the non-Gaussian parameter. Relaxation
dynamics of the larger particles varies with $A$ and the distance from 
the barrier. In contrast to the smaller particles, the relaxation of 
$F_{s,z}(q,t)$ of the larger particles is very slow. Interestingly, the
relaxation of $F_{s,z}(q,t)$ for the smaller particles is nearly five
times faster then the larger particles at both temperatures, shown 
in Fig. \ref{f:fszkt}. This faster density relaxation of the smaller 
particles is because it is measured at the wave number associated with
the first peak of the static structure factor. This wave number is 
corresponding to the inter-particle separation. Since the volume fraction 
is low and the smaller particles are mostly away from the barrier, thus 
they move faster at the length scale of the inter-particle separation. 
On contrary, the smaller particles exhibit the slower density relaxation,
measured at the length scale of the simulation box ($q=2\pi/L_z$), as 
shown by Kumar \cite{j:anil_sym_jcpdyn}. Because the smaller particles 
are trapped between two potential barriers separated by the box length 
$L$, due to the periodic boundary conditions. At $T=$ 0.5, the dynamics 
of the larger particles is slower near the barrier compared to far from 
the barrier. Interestingly, at $T=$ 0.3, $F_{s,z}(q,t)$ near the barrier
start showing a very little hump, which enhances with increasing
asymmetry parameter $A$. A presence of the hump signifies the cagelike 
motion of the larger particles near the barrier, which support the 
behavior of the non-Gaussian parameter and the self part of the van 
Hove correlation along the $z-$direction. For a direct comparison of 
$F_{s,z}(q,t)$ at $T=$ 0.3, we plot it in Fig. \ref{f:fszktl}, at
$z=$ 8.0 near the asymmetric barrier. We then calculate the 
$\alpha-$relaxation time, $\tau_\alpha$ and fitted
the $F_{s,z=8.0}(q,t)$ with the KWW function to get exponent $\beta$.
Here $\tau_\alpha$ is defined as $F_{s,z}(q,t=\tau_\alpha)=e^{-1}$.
Left bottom inset of Fig. \ref{f:fszktl} shows that $\tau_\alpha$ 
increases (though little) with the asymmetry parameter $A$. The KWW 
exponent $\beta$ is around 0.96 at $A=$ 12, which entails the presence
of non-exponential density relaxation near the barrier for the larger
particles; the tail of $F_{s,z=8.0}(q,t)$ is extended.

\section{Summary and conclusions\label{s:sumcon}}

We have shown the effect of asymmetry in the external potential barrier along the
$z-$direction, on the structure and dynamics of the binary colloidal mixture using
molecular dynamics simulations. Due to the depletion interaction between the potential
barrier and the larger particles, the density of the larger particles increases near 
the barrier. However, as the asymmetry in the potential barrier increases, barrier 
crossing becomes less probable, leading to even the higher densities near the barrier
especially on the asymmetric side. This leads to transient caging of the larger particles.
Therefore, for highly asymmetric potential, the diffusion of the larger particles crosses
over to super-Arrhenius from sub-Arrhenius diffusion. In general, depletion interactions 
between the potential barrier and the larger particles makes the activation energy to 
jump over the barrier, temperature dependent. However, the nature of this dependency
changes with the asymmetry parameter in the external potential. At low values of $A$,
the activation energy of the larger particles decreases with temperature, whereas for
the larger values of $A$, it increases with temperature. Thus, the dynamics of the 
larger particles at very large values of $A$ is very similar to the dynamics of
supercooled liquids, where super-Arrhenius diffusion is
observed \cite{wales1,wales2,ysingh}. We expect that at the larger asymmetric 
potential barrier heights, the super-Arrhenius diffusion will enhance and
subsequently the deformation parameter will further increase. This will be 
interesting since the volume fraction we have used in this study is very low 
compared to that of supercooled liquids. 
Moreover, asymmetric potentials are used in the transport of molecules  
in biological channels \cite{shaw,chinappi,comelles}, targeted delivery of colloids
by bacteria \cite{koumakis}, and Brownian motors \cite{rosenbaum,ai}. Most of these
studies are carried out for single-component systems in an asymmetric potential
subjected to various conditions of driving. It will be interesting to investigate 
the dynamics of binary mixtures in these scenarios and the system we simulated 
provides a model for these investigations. Finally, we expect that the 
experimental realization of our model is straight forward. Spatially asymmetric
periodic potentials have been realized in experiments via dielectrophoretically induced 
forces \cite{rousselet,gorre-talini} as well as using optical tweezers \cite{thorn}. 
Dynamics of single-component colloidal systems subjected to an external sinusoidal 
potential has been studied experimentally by Dalle-Ferrier \textit{et al.} \cite{egelhaaf}.
They have carried out these investigations on two different colloidal systems whose
size ratio is close to 1:2, which has been used in our simulations.

\begin{acknowledgments}
	The authors acknowledge financial support from the Department of Atomic Energy, India 
	through the 12th plan project (12-R\&D-NIS-5.02-0100). The data that support the findings 
	of this study are available from the corresponding author upon reasonable request.
\end{acknowledgments}

\appendix

\renewcommand{\thefigure}{A.\arabic{figure}}

\section{Simulation details\label{sa:simdet}} 
\begin{figure}
	\includegraphics[width=8cm, height=6.5cm]{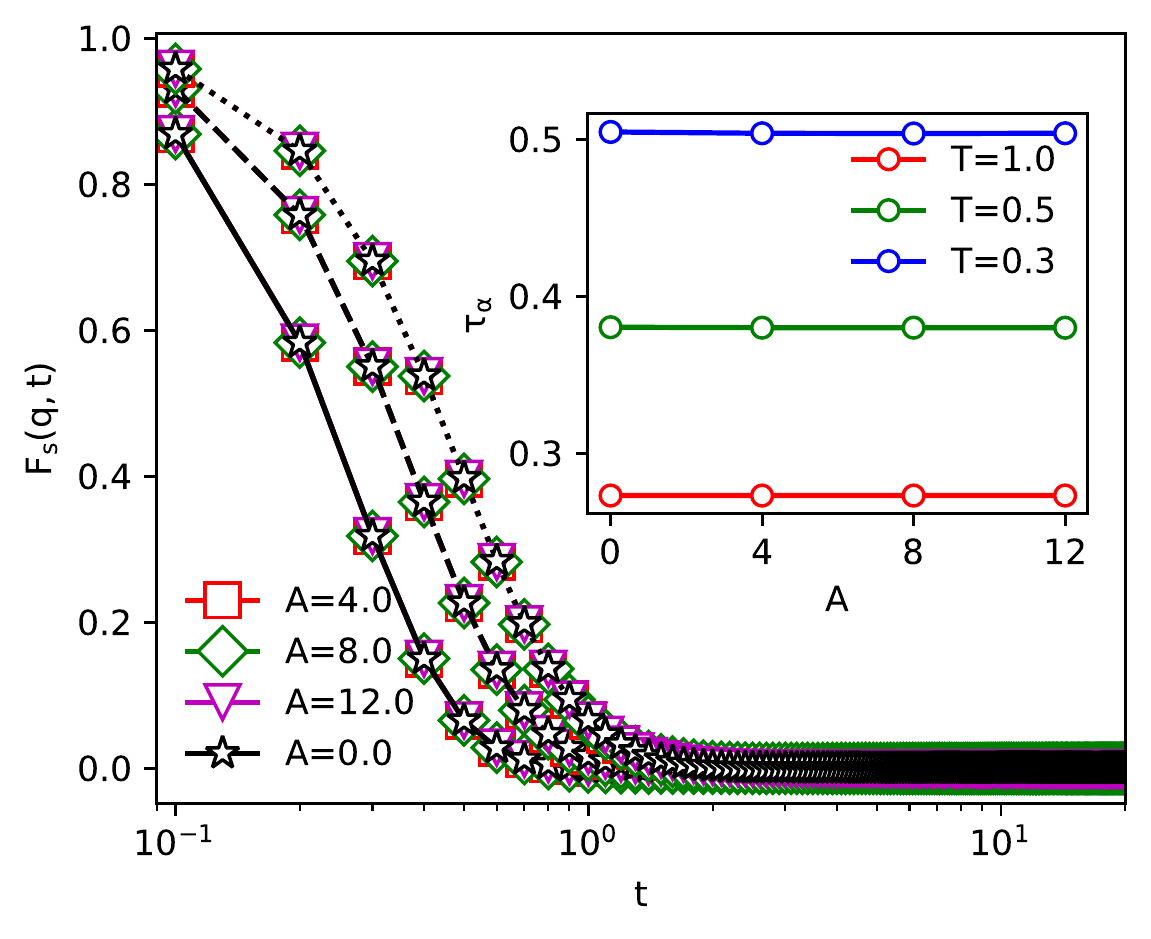}
	\caption{\label{f:eqfskt} Incoherent intermediate scattering function,
	$F_s(q,t)=N^{-1}\langle \sum_{i=1}^N e^{-i\,{\mathbf q}.{\mathbf r}_i} \rangle$
	is plotted against time $t$. Here, $q$ is the wave number corresponding to the 
	main (first) peak of the static structure factor $S(q)$ and $N$ is the total
	number of particles in the system. Solid, dahsed, and 
	dotted lines are corresponding to $T=$ 1.0, 0.5, and 0.3, respectively. 
	The $\alpha-$relaxation time increases with lowering temperature, however,
	it is invariant with $A$; at the same temperature, $F_s(q,t)$ coincides for
	different $A$. At the lowest temperature, i.e., $T=$ 0.3,
	$\tau_\alpha \simeq$ 0.5, and the equilibration time used is 1000$\tau_{LJ}$
	that is much longer than 100$\tau_\alpha$.}
\end{figure}

In this equivolume sharing binary mixture at volume fraction $\phi=$ 0.2, the number 
of large and small particles are 117 and 938, respectively at $L=$ 17. Thus, the total 
number of particles in the system are $N=$ 1055. The system is simulated in constant 
NVT ensemble, where the equations of motion, $\dot{{\mathbf q}}_i = \mathbf p_i/m$ 
and $\dot{{\mathbf p}}_i = \mathbf F_i - \lambda \, \mathbf p_i$, are simultaneously
integrated using the fifth-order Gear predictor-corrector method \cite{ham} at the
time step of $dt=$ 0.001. Here, $\mathbf q_i$, $\mathbf p_i$, and $\mathbf F_i$ are 
position, momentum, and force acting on an i$th$ particle, respectively. The parameter
$\lambda$ is the thermostat factor governed by the Gaussian principle of least 
constraint \cite{EvansMorris}. The system is equilibrated at $T=$ 3.0 for 
$1\times10^6$ steps, then this equilibrated configuration is used as the initial 
configuration for the rest of the temperatures at the corresponding $A$. At temperatures
$T=$ 2.0--0.75, the system is equilibrated till $1\times10^6$ steps, followed 
by $4\times10^6$ production runs. However, at low temperatures, i.e., $T=$ 0.5 onwards, 
the system is equilibrated till $1\times10^7$ steps, followed by $1.5\times10^7$ 
production runs. The equilibration times used in this study are much longer than 
100$\tau_\alpha$ at all temperatures and asymmetry 
parameters (see Fig. \ref{f:eqfskt}). To look at the finite-size effects, we 
simulated the system at box lengths $L=$ 15, 19, and 21, as the equivolume mixtures.
The number of particles at these box lengths are $N=$ 724, 1472, and 1989, respectively.
Note that the simulation parameters, i.e., time step, equilibration and production
runs, etc., are same as the $L=$ 17. We used the same simulation procedure for 
all independent runs and the different box lengths.

\section{Finite-size effects\label{sa:finsize}}
\begin{figure}
	\includegraphics[width=7cm, height=10.5cm]{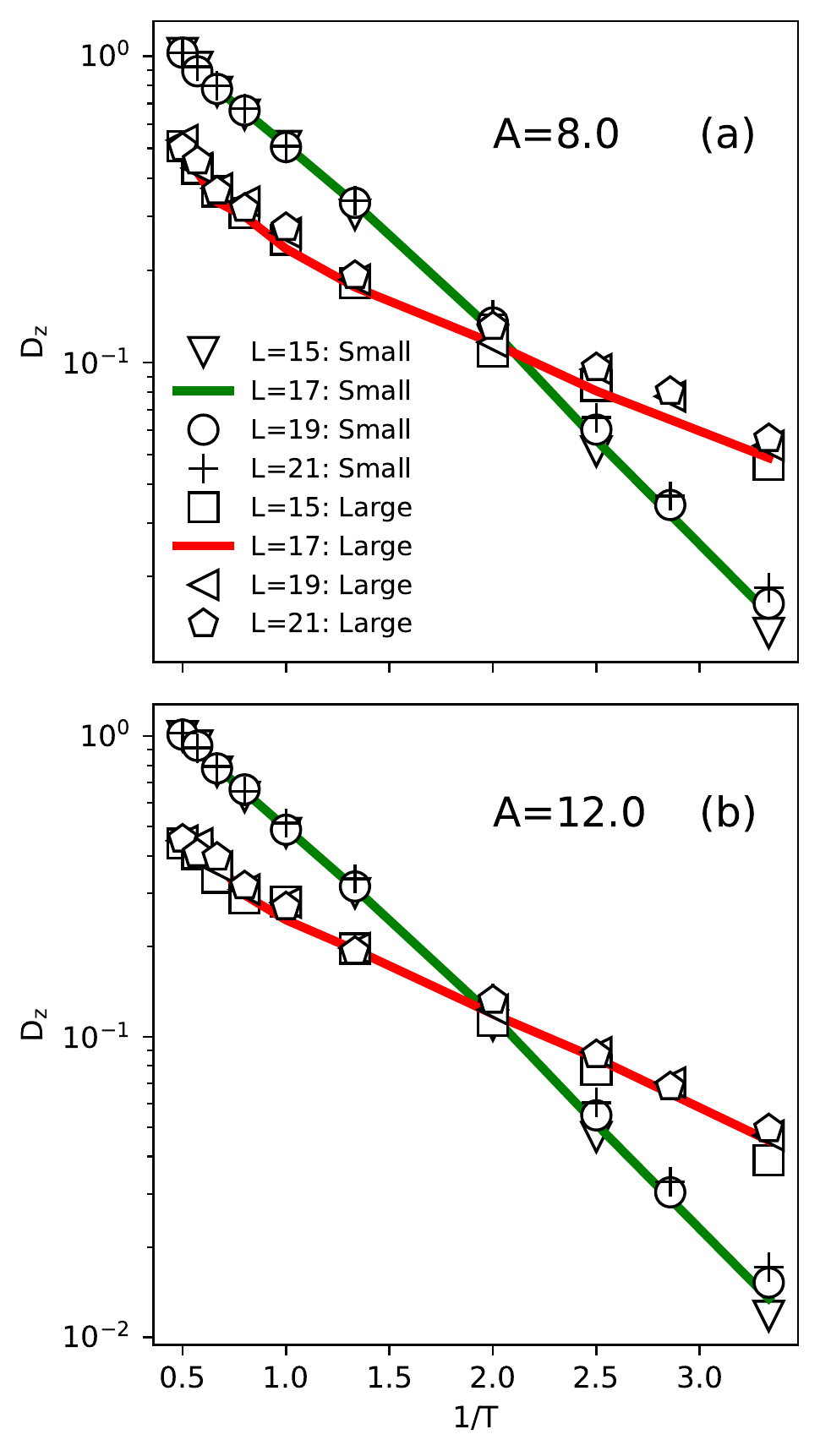}
	\caption{\label{f:dcmL} A comparison of diffusion at four different box 
	lengths, $L=$ 15, 17, 19, and 21 at the asymmetry parameters $A=$ 8.0 and 12.0.
	The legend of (a) is applied to (b).}
\end{figure}

We calculate the diffusion coefficient of large and small particles at $L=$ 15, 19,
and 21, and compare them with the results at $L=$ 17, to examine finite-size
effects in the system (see Fig. \ref{f:dcmL}). When the barrier is symmetric,
the dynamics is unaffected by changing the box lengths, as shown by 
Mustakim \textit{et al.} \cite{j:anil_soft_sym}. Therefore, we compare diffusion 
of small and large particles at the symmetry parameters $A=$ 8.0 and 12.0, where the
variations in the density profile of the larger particles with $A$ are 
pronounced [see Figs. \ref{f:dfl}(a--b)]. Figure \ref{f:dcmL} shows that diffusion 
of the smaller and the larger particles is (approximately) invariant with varying
the box length. We change the length of the cubical box and apply external potential 
at the middle of the z-axis, which adds a periodicity of length $L$ due to the PBCs,
therefore, the finite-size effects are unexpected. Thus, the finite-size effects in
our simulations of the binary colloidal mixture subjected to the external potential
barrier, are insignificant, at the given simulation parameters.
\begin{figure*}
	\includegraphics[width=14cm, height=10cm]{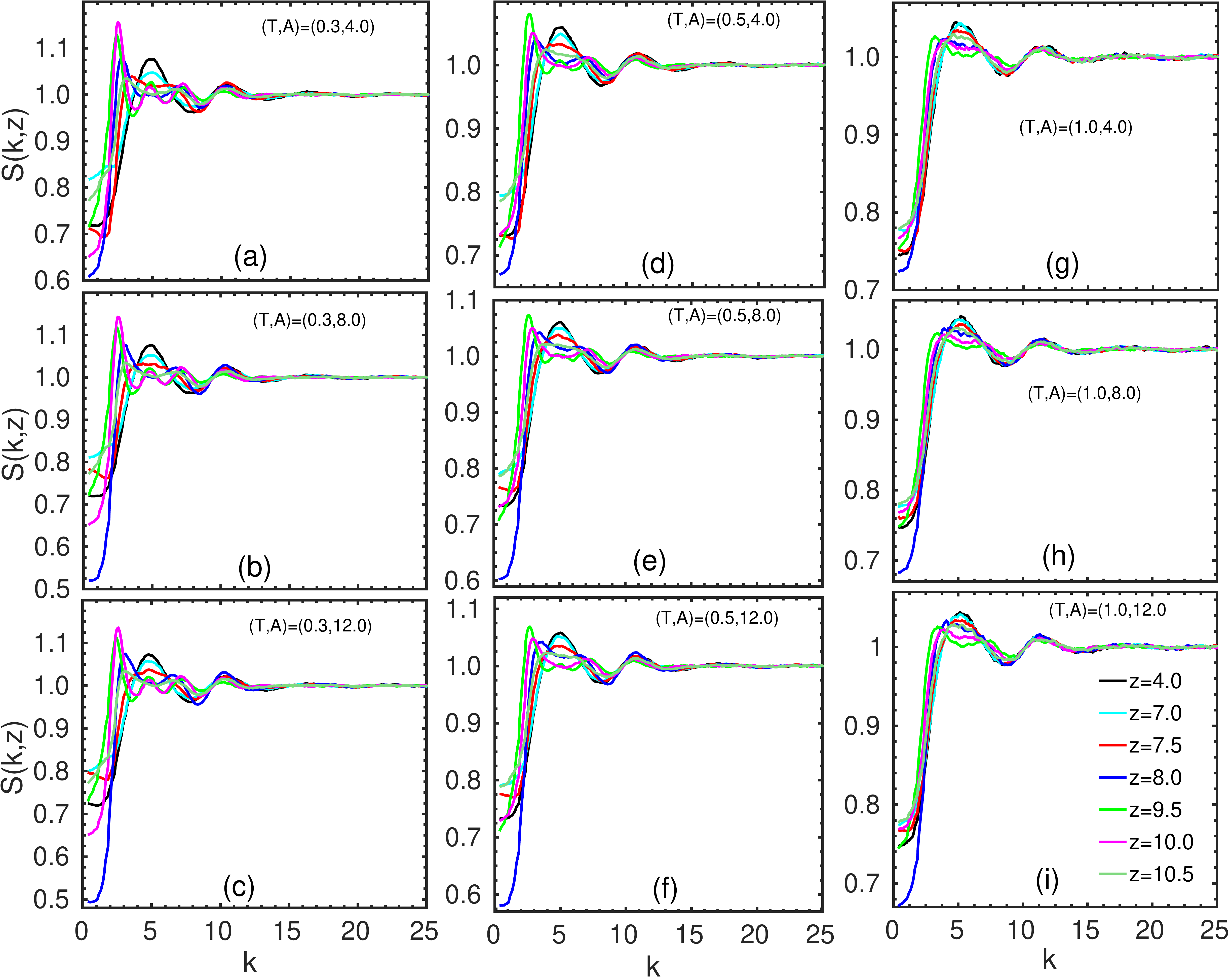}
	\caption{\label{f:skz} Two-dimensional structure factor near ($z=$ 7.0, 7.5, 8.0, 
	9.5, 10.0, 10.5) and far ($z=$ 4.0) from the external barrier at 
	asymmetry parameter $A=$ 4.0, 8.0, 12.0. (a--c) $T=$ 0.3, (d--f) $T=$ 0.5,
	and (g--i) $T=$ 1.0. Legend of (i) is applied to (a--h).}
\end{figure*}
\begin{figure}
	\includegraphics[width=8cm, height=9cm]{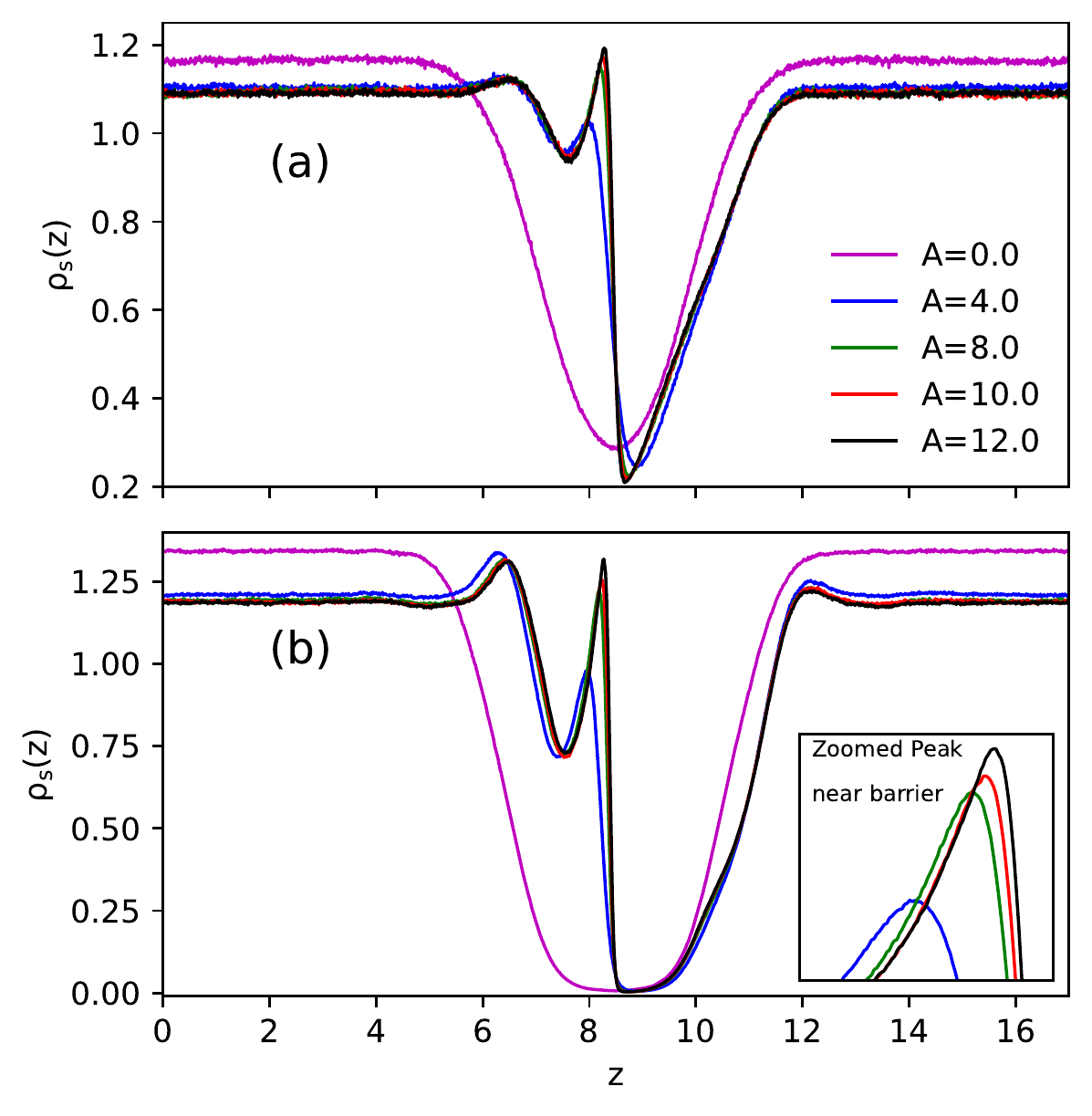}
	\caption{\label{f:dfs} Density profile of the smaller particles at
	$A=$ 0.0, 4.0, 8.0, 10.0, and 12.0. (a) $T=$ 1.0 and (b) $T=$ 0.3.
	The zoomed area of the peak, near the barrier at $T=$ 0.3, is shown in
	the inset of (b), showing an increment in its height with position
	shifting towards right. This signifies a little mixing of the smaller 
	particles towards the asymmetric side of the barrier.}
\end{figure}
\begin{figure}
	\includegraphics[width=8cm, height=9cm]{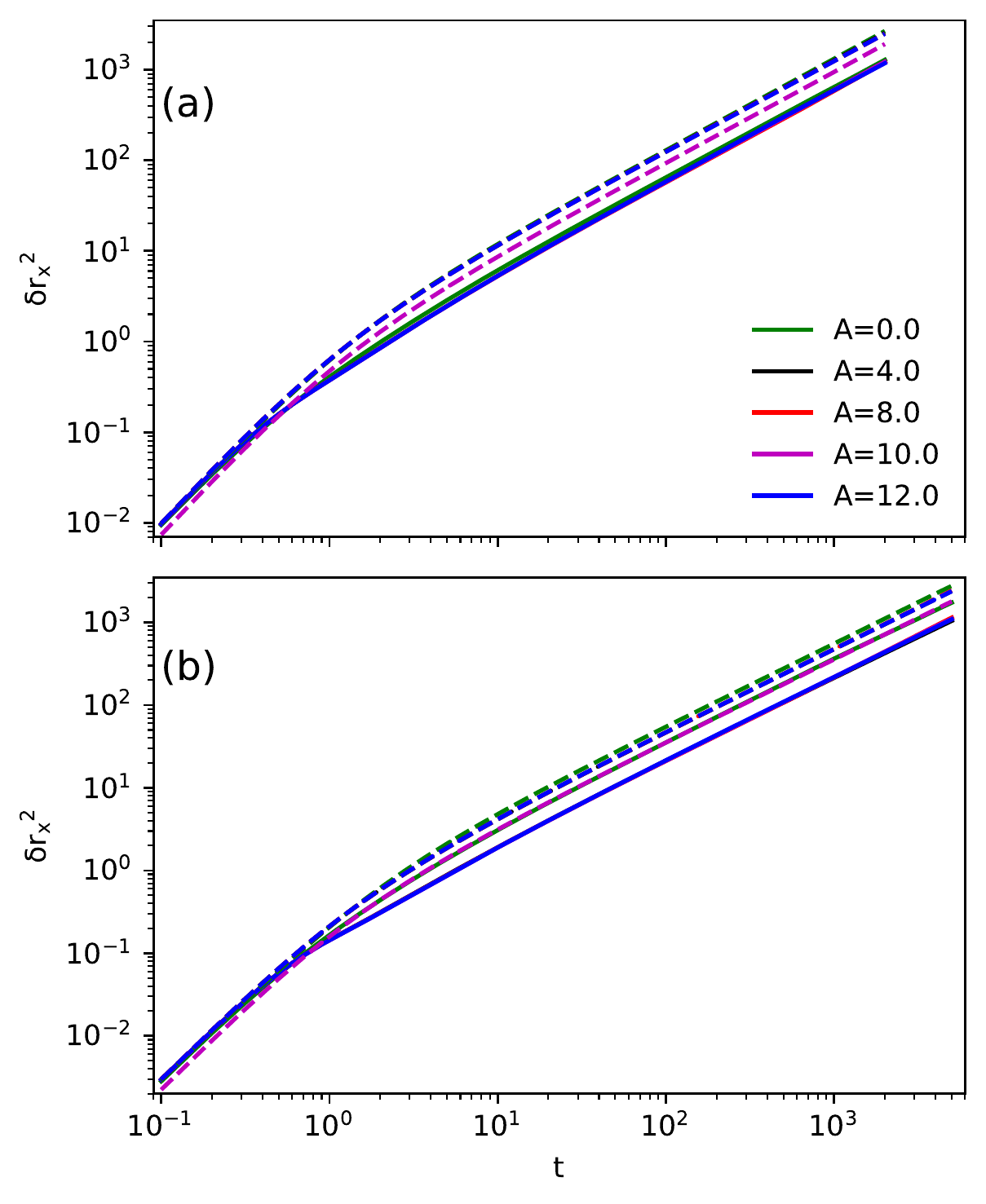}
	\caption{\label{f:msdx} Mean-squared displacement of the 
	smaller (dashed lines) and the larger (solid lines) particles
	along X-direction at $A=$ 0.0, 4.0, 8.0, 10.0, and 12.0.
	(a) $T=$ 1.0 and (b) $T=$ 0.3.}
\end{figure}
\section{Structure factor\label{sa:sk}}
To look at whether the local freezing of the larger particles into layers in 
$xy-$plane is taking place along the $z-$direction, we calculate $z-$dependent 
two dimensional (2D) structure factor (near and far from the barrier), which
is defined as 
\begin{equation}
	S(k,z) = \frac{1}{N(z)} \sum_{i,j=1, \, j \ne i}^{N(z)} 
	\exp[-i\{\mathbf k.(\boldsymbol \xi_i(z)-\boldsymbol \xi_j(z))\}] ,
\end{equation}
where $\boldsymbol \xi_i(z) = x_i \hat x + y_i \hat y.$, $N(z)$ is the
total (small and large) number of particles in a 2D layer at z, and wave vectors
are generated in 2D plane. We slice the simulation box along $z-$direction with a 
width $dz=$ 0.5 to calculate $S(k,z)$, which is shown in Figs. \ref{f:skz}(a-i). 
$S(k,z)$ is calculated at three representative temperatures, \textit{viz}, 
$T=$ 1.0, 0.5, 0.3 at asymmetry parameters $A=$ 4.0, 8.0, 12.0.
At $T=$ 1.0 [see Figs. \ref{f:skz}(g-i)], the 
binary colloidal mixture shows normal liquid structure, as $S(k,z)$ shows a single 
broad peak at $k\simeq$ 5, away from the barrier, and $S(k,z)$ split in to two peaks
around $k\simeq$ 3.0 and 6.0, near the barrier. These peaks are corresponding to the two
length scales ($k=2\pi/\sigma_{ss,ll}$) due to two different sizes of the particles.
At this temperature, $S(k,z)$ slightly differs at different $A$, specially near the 
barrier, which is usual because the density profile is different as discussed in the 
main text. The non-zero values of $S(k,z)$ at low wavenumbers shows the demixing of
the particles in the system. At $T=$ 0.5 [see Figs. \ref{f:skz}(d-f)], $S(k,z)$ shows
pronounced splitting of its main peak near the barrier along with multiple secondary 
peaks. These peaks are corresponding to the multiple secondary structures near the 
barrier due to crowding, as seen in the density profile also. At the lowest 
temperature of the study, i.e., $T=$ 0.3 [see Figs. \ref{f:skz}(a-c)], the first peak 
height of $S(k,z)$ further increases with more splitting in the secondary peak, near
the barrier. These secondary peaks are due to multiple secondary structures near the 
barrier, which is pronounced at low temperatures. Thus, $S(k,z)$ does not show any 
signature of crystallization of layers near the barrier, although local crowding
is present.

\section{Smaller particles' density profile \label{sa:dfs}}
The dynamics of the smaller particles, discussed in the main paper, shows the
anomaly (along $z-$direction) at the asymmetry parameter $A=$ 10. The diffusion 
coefficient decreases and the activation energy increases at $A=$ 10. To examine
this, we calculate the density profile of the smaller particles, along the 
direction of applied external potential at one high ($T=$ 1.0) and one low 
temperature ($T=$ 0.3), which is shown in Figs. \ref{f:dfs}(a--b). As $A$ 
increases, (small) peaks in $\rho_s(z)$ develop near the asymmetric side of the
barrier, which shows a little attraction of the smaller particles also. However, 
their number near the barrier is very small compared to the larger particles.
This attraction of the smaller particles towards the asymmetric side of the 
barrier enhances (slightly) at low temperatures and the higher asymmetries. The zoomed 
peak in the inset of Fig. \ref{f:dfs}(b) clearly shows a little increment and
shift in the peak near the barrier. Also, a little hump appears towards the 
symmetric side of the barrier at $T=$ 0.3, for $A=$ 4 and above. This contrasting nature
of $\rho_s(z)$ compared to the symmetric barrier shows that the enhanced 
activation energy and the smaller diffusion coefficient (at $A=$ 10) is 
because of a very small number of the smaller particles near the barrier.
Actually, the number of small particles is very small near the barrier 
to show any characteristic of the cagelike motion.

\section{Dynamics along $X-$direction\label{sa:msdx}}
In a study of the binary colloidal mixture, subjected to symmetric external
potential barrier along one of the spatial direction, shows that dynamics 
of both types of particles along $X-$ and $Y-$directions is not affected by the
barrier \cite{j:anil_sym_jcpdyn}. To look at the effect of the asymmetry on
the dynamics of the particles in the mixture along one of the directions
normal to the $z-$direction, we plot the MSD's of (both types of) particles at 
$T=$ 1.0 and 0.3. Figs. \ref{f:msdx}(a-b) show that $\delta r^2_x$ of the 
larger particles is invariant with $A$ at $T=$ 1.0, however, it slows down 
slightly at $A\ne$ 0 and $T=$ 0.3. The smaller particles' MSD along $X-$ direction 
also shows a little slow down at $A=$ 10.

\bibliography{colloid} 

\end{document}